\pdfoutput=1

\documentclass[11pt]{article}
\usepackage{graphicx}
\usepackage[final]{acl}

\usepackage{times}
\usepackage{latexsym}
\usepackage{mdframed}
\usepackage{dashrule} 
\usepackage{booktabs}
\usepackage{cuted}
\usepackage{tabularx} 
\usepackage{ragged2e} 
\usepackage{multirow}
\usepackage{graphicx}
\usepackage{tcolorbox}
\usepackage{tabularx}   
\usepackage{array}
\usepackage{amsmath}
\usepackage{tikz}
\usepackage{amsmath}
\usepackage{makecell}
\usepackage{listings}
\usepackage{amsmath}
\usepackage{mdframed}
\usepackage{booktabs}
\usepackage[table]{xcolor}
\usepackage[linesnumbered, ruled, vlined, noend]{algorithm2e}

\newmdenv[
  linecolor=black,
  linewidth=0.5pt,
  leftmargin=0pt,
  rightmargin=0pt,
  skipabove=6pt,
  skipbelow=6pt,
  innerleftmargin=6pt,
  innerrightmargin=6pt,
  innertopmargin=6pt,
  innerbottommargin=6pt,
  roundcorner=0pt,
  topline=true,
  bottomline=true,
  rightline=true,
  leftline=true,
  linestyle=dashed
]{dashedbox}
\lstdefinestyle{demo}{
    basicstyle=\fontsize{8}{10}\ttfamily,
    keywordstyle=\color{blue},
    commentstyle=\color{gray},
    stringstyle=\color{green},
    showstringspaces=false,
    breaklines=true,
    breakatwhitespace=false,
    breakindent=0pt,
    escapeinside={(*@}{@*)},
    literate={á}{{\'a}}1 {ã}{{\~a}}1 {é}{{\'e}}1,
}
\usepackage{amssymb}
\usepackage{fvextra}

\usepackage[T1]{fontenc}

\usepackage[utf8]{inputenc}

\usepackage{microtype}

\usepackage{enumitem}
\newcommand{\ours}{\textsc{Dreams}}
\newcommand{\se}{$\text{FGE}^2$}
\newcommand{\re}{$\text{PE}^2$}
\newcommand{\ratio}{$\text{CGE}^2$}

\title{Towards Effective Structured Context Modeling for Conversational Recommender Systems via Dual-node Monte Carlo Tree Search}

\author{
Jincheng Zhang$^{1,4}$\quad Chen Huang$^{1,2,4}$\thanks{Corresponding author.}\quad
\textbf{Wenqiang Lei}$^{1,4}$ \quad \textbf{See-Kiong Ng}$^{2}$ \quad Yang Deng$^{3}$\\
$^{1}$ College of Computer Science, Sichuan University  \\
$^{2}$ Institute of Data Science, National University of Singapore   \\
$^{3}$ School of Computing and Information Systems, Singapore Management University\\
$^{4}$ Engineering Research Center of Machine Learning and Industry Intelligence, \\Ministry of Education, China \\
\texttt{\{erzmuxin, huangc.scu\}@gmail.com}
}

\begin{document}
\maketitle
\begin{abstract}
We investigate the role of conversational context modeling in user preference tracking for Conversational Recommendation Systems (CRSs). In this regard, we propose \ours, a novel tree-structured context modeling framework that explicitly captures user preference evolution throughout multi-turn interactions. \ours\ introduces two specialized node types to support the two fundamental objectives of CRSs: preference elicitation and preference exploitation. Specifically, elicitation nodes leverage Monte Carlo Tree Search (MCTS) to strategically explore conversational actions and infer latent user preferences, while exploitation nodes employ LLM-based refinement to transform the tracked preference state into structured retrieval queries for recommendation. Extensive experiments on benchmark datasets demonstrate the effectiveness of \ours\ and its design. Our code is available at \url{https://github.com/SCUNLP/DREAMS}.
\end{abstract}

\section{Introduction}
\begin{figure}
    \centering
    \includegraphics[width=1.0\linewidth]{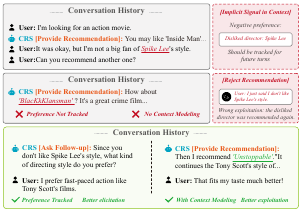}
    \caption{The claim `\textit{not a big fan of Spike Lee’s style}' expresses a negative preference that should be retained across turns. Otherwise, the CRS may continue recommending his films, leading to repeated user rejection.
    }
    \label{fig:intro}
\end{figure}

\noindent Conversational Recommendation Systems (CRSs) aim to actively elicit user preferences through multi-turn conversations (i.e., \textit{preference elicitation}) and leverage the acquired preferences to recommend relevant items (\textit{preference exploitation}) \cite{deng2021unified, lei2020estimation, he2023large, deng2025proactive}. Achieving these goals requires accurate user preference tracking (i.e., what the user likes and dislikes), as insufficient tracking can result in redundant questions or premature and irrelevant recommendations, thereby degrading system effectiveness. This, in turn, necessitates robust conversational context modeling \cite{chen2025recusersim, 10.1145/3726302.3730080}, since such user preferences are progressively revealed throughout the conversation. Taking Figure \ref{fig:intro} for example, the conversation history may implicitly reveal a negative preference toward a specific director. Such implicit signals should be tracked and consistently suppressed in future recommendations involving that director. Therefore, conversation context modeling forms the foundation of preference tracking, which ultimately supports both preference elicitation and preference exploitation.

However, existing methods frequently model conversational context as simple free-form text for user preference tracking, which leads to suboptimal performance \cite{chen2024all, huang2024concept} due to two limitations. 
1) For preference elicitation, these methods directly inject the entire conversation history into LLM prompts to decide whether to ask or recommend \cite{feng2023large, fang2024multi, wang2023rethinking, qin2024beyond}, resulting in information overload and untargeted questioning \cite{kemper2024retrieval, zhang2023vague}. 2) For preference exploitation, they encode free-form conversational context (as a query) into dense embeddings for item retrieval, where irrelevant conversational content introduces noise and degrades retrieval accuracy \cite{li-etal-2025-graphotter, ni2025trustworthyretrievalaugmentedgeneration, xu-etal-2025-beyond}. To ease these limitations, recent studies have represented conversations as JSON-formatted dialogue states \cite{kemper2024retrieval} or MCTS \cite{du-etal-2025-sapient, dao-etal-2024-experience}. 
However, JSON-based methods represent each turn as an isolated state snapshot, overlooking the dependencies among preference states across the conversation. Conversely, MCTS-based methods capture the exploration structure of interactions, but their nodes lack explicit semantic representations of user preferences. What's more, they only partially structures preference elicitation, while not to jointly structure both elicitation and exploitation under a unified framework.
As a result, neither paradigm simultaneously enables effective preference tracking.

To validate the above analysis, we conducted a targeted evaluation of existing CRSs in Section \ref{evaluation_protocol}. The results conclusively demonstrate that existing methods conclusively hinder user preference tracking, diminishing both preference elicitation and exploitation, and thus overall system performance. This is evidenced by their difficulty in identifying and utilizing conversation state for accurately timing questions or recommendations. Furthermore, these CRSs exhibit high recommendation error rates, even when the conversation context contains complete user preferences. These findings emphasize the need for advanced context modeling to effectively track user preferences and suggesting that effective user preference tracking is not merely a matter of structuring dialogue history. It requires the structured state to actively drive both preference elicitation and exploitation.

To this end, we propose \ours, a \underline{\textbf{D}}ual-node conversational \underline{\textbf{R}}ecomm\underline{\textbf{E}}nd\underline{\textbf{A}}tion system using \underline{\textbf{M}}onte carlo tree \underline{\textbf{S}}earch for effective user preference tracking. 
In particular, \ours\ explicitly models user preference evolution through a tree-structured conversation state, representing conversation states as Json-structured nodes and state transitions as edges, forming an evolving preference structure throughout the conversation.
Importantly, \ours\ introduces two complementary node types. Elicitation Nodes focus on acquiring user preference by strategically exploring alternative conversational actions (e.g., \textit{DirectorInquiry, FailureReflection}) through Monte Carlo Tree Search (MCTS), enabling the system to infer and refine latent user preferences. Exploitation Nodes focus on recommendation by transforming the tracked preference state into structured retrieval queries, thereby reducing contextual noise and improving item retrieval accuracy. 
The two node types are coupled through the same evolving preference state: ELNodes update the state through elicitation and feedback, whereas EXNodes use that state for retrieval, after which recommendation feedback is written back for subsequent decisions.
As such, \ours~provides a structured representation of the entire user preference lifecycle, cohesively bridging conversation flow and recommendation queries to optimize the elicitation and exploitation of user information for superior recommendations.

We experimentally demonstrate the effectiveness of \ours~using the widely-used simulator-based evaluation framework\footnote{Human evaluation in Section \ref{in-depth} shows our reliability.} \cite{wang2023rethinking, huang2024concept} and two benchmark datasets. The results show that \ours~achieves superior recommendation accuracy compared to existing methods (+7.43\%, on average), exhibiting enhanced performance in both preference elicitation (+9.35\%) and exploitation (+4.00\%). 
Further analysis reveals that our structured context modeling not only provides fine-grained and interpretable guidance for conversational planning, but also enables the CRS to accurately associate subtle or implicit contextual signals with the corresponding user preference attributes through structured representations and search over conversational states. In summary, our contributions are as follows:

\begin{itemize}[leftmargin=*]

    \item We emphasize the importance of structured context modeling for effective preference tracking and systematically analyze the limitations of existing LLM-based CRS methods in this regard.

    \item We propose \ours, a unified framework that jointly models preference elicitation and exploitation through MCTS over structured conversational states.

    \item Extensive experiments demonstrate the effectiveness of \ours~and validate the critical role of structured context modeling in conversational recommendation.

\end{itemize}

\section{Related Work}
\label{sec:related_work}

\noindent\textbf{User Preference Tracking}. 
A central challenge in conversational recommendation is maintaining an accurate representation of user preferences as they are progressively revealed throughout multi-turn interactions \cite{liu2023conversational, deng2025proactive}. Existing LLM-based CRSs primarily support this process from two perspectives. For preference elicitation, prior work focuses on improving dialogue decision-making, such as determining when to ask questions versus provide recommendations \cite{gao2023chat, friedman2023leveraging, wang2023rethinking}. For preference exploitation, most methods encode conversational history into semantic representations for downstream retrieval and recommendation \cite{gao2023chat, liu2023conversational, huang2024concept, qin2024beyond}.  
Our work highlights the importance of structured context modeling for maintaining an evolving preference representation that can simultaneously support both preference elicitation and preference exploitation.

\noindent\textbf{Context Modeling in LLM-based CRS}.
Most LLM-based CRSs rely on free-form dialogue history as the underlying representation of user preferences, using it for both preference elicitation \cite{gao2023chat, friedman2023leveraging, wang2023rethinking} and preference exploitation \cite{he2023large, qin2024beyond}. Such unstructured context can obscure preference signals and introduce retrieval noise \cite{zhang2023vague, kemper2024retrieval, li-etal-2025-graphotter, huang2024concept, ni2025trustworthyretrievalaugmentedgeneration}, even for long-context LLMs\footnote{Also evidenced by our experiment in Table \ref{tab:gemini}.} \cite{liu2023lost, bai2023longbench, an2025beyond}. 
Recent work attempts to alleviate this issue from two perspectives. RA-CRS \cite{kemper2024retrieval} summarizes dialogue history into JSON-formatted preference states, but does not model preference state evolution during conversational exploration. In contrast, MCTS-based CRSs such as SAPIENT \cite{du2024sapient} and T-EPL \cite{dao-etal-2024-experience} explicitly model exploration through tree search, yet their search nodes are not grounded in structured preference states. Consequently, existing methods either provide structured preference states without structured preference evolution, or model exploration trajectories without structured preference states\footnote{Refer to Appendix \ref{app:comparison} for method comparison.}. Motivated by this gap, our framework performs search directly over structured conversational states to jointly support preference elicitation and exploitation.

\section{Preliminary Study}
\label{sec_pre}
We conduct a targeted evaluation to analyze the deficiencies of existing LLM-based CRSs in effective context modeling, leading to challenges in preference elicitation and exploitation. More details are provided in Appendix \ref{app_metric}, \ref{app_evaluator}.

\subsection{Evaluation Setup}\label{evaluation_protocol}
\noindent\textbf{Evaluation Overview.} Adhering to established practices \cite{wang2023rethinking, wang2023improving, fang2024multi, qin2024beyond, huang2024concept}, we employ LLM-based user simulators to interact with CRSs. The interaction continues until either the simulator accepted a recommendation or the maximum allowed number of turns is reached. Each resulting conversational history, as the context, is subsequently utilized for evaluation purposes. See Appendix \ref{app_user} for details.

\begin{table}[t]
    \centering
    \resizebox{0.49\textwidth}{!}{
    \begin{tabular}{l c c c c}
        \toprule
        \multirow{2}{*}{Model} & Accuracy & \multicolumn{2}{c}{Elicitation} & Exploitation \\
        \cmidrule(lr){3-4}
        & SR$\uparrow$ & \se$\downarrow$ & \ratio$\downarrow$ & \re$\downarrow$ \\
        \midrule
        InterCRS 
        & 0.313 
        & 0.360 
        & 0.570
        & 0.240 \\
        \midrule
        MACRS 
        & 0.400
        & 0.187 
        & 0.456
        & 0.220 \\
        \midrule
        RA-CRS 
        & 0.333 
        & 0.333
        & 0.461
        & 0.240 \\
        \midrule
        ChatCRS
        &\textbf{0.440} 
        & \textbf{0.120}
        & 0.561
        & \textbf{0.200} \\
        \midrule
        PC-CRS
        & 0.403
        & 0.247
        & 0.491
        & 0.240 \\
        \midrule
        SAPIENT-LLM
        & 0.380 
        & 0.307
        & \textbf{0.443}
        & 0.233 
        \\
        \midrule
        T-EPL
        & 0.327 
        & 0.313
        & 0.469
        & 0.253 
        \\
        \bottomrule
    \end{tabular}
    }
    \caption{Elicitation and exploitation error evaluation. }
    \label{tab:ep_results}
\end{table}

\noindent\textbf{Datasets \& User Simulator}. Following prior CRS evaluation protocols \cite{wang2023rethinking, huang2024concept, qin2024beyond}, we use LLM-based user simulators with persona-conditioned preference profiles (e.g., genres, actors, and directors). Experiments are conducted on Redial \cite{li2018towards}\footnote{Additional datasets are introduced in Section \ref{main_settings}}. More details are provided in Appendix \ref{app_user}. Complementary human validation is reported in Section \ref{main_results} and Appendix \ref{app_human}.
More details are provided in Appendix \ref{app_user}.

\noindent\textbf{Baselines}. Our preliminary study incorporates free-form-based LLM-based CRSs, including InterCRS\cite{wang2023rethinking}, MACRS\cite{fang2024multi}, ChatCRS\cite{li2024incorporating}, PC-CRS\cite{qin2024beyond}, and RA-CRS\cite{kemper2024retrieval}. Additionally, we add two MCTS-based baselines, SAPIENT \cite{du-etal-2025-sapient} and T-EPL \cite{dao-etal-2024-experience}. See Appendix \ref{app_baselines} for details.

\noindent\textbf{Evaluation Metrics.} 
We introduce specific error-based metrics to investigate the deficiencies in preference elicitation and exploitation\footnote{Detailed in Table \ref{tab:ep_metrics} in Appendix.}. (1) For preference elicitation, we consider two key metrics: \underline{Coarse-Grained Elicitation Error} (\ratio) and \underline{Fine-Grained Elicitation Error} (\se). The former quantifies the proportion of wrong action selection regarding asking question and making recommendations, while the latter quantifies the proportion of ineffective or misaligned follow-up questions after the system receives specific user feedback about their preferences. (2) To assess preference exploitation, we employ the \underline{Preference Exploitation Error} (\re), which measures the proportion of instances where the CRS fails to retrieve a relevant item for recommendation despite a fully clarified understanding of user preferences. (3) Furthermore, in accordance with established methodologies \cite{qin2024beyond, fang2024multi}, we utilize the conventional metrics of \underline{Success Rate} (SR) to assess the overall recommendation success. Refer to Appendix \ref{app_metric} for implementation details and Appendix \ref{app_human} for reliability analysis regarding LLM-as-judge.

\subsection{Experimental Findings}
Table \ref{tab:ep_results} shows that off-the-shelf CRS still suffer from persistent elicitation and exploitation errors, highlighting the need for more faithful and fine-grained context modeling. In preference elicitation, coarse-grained action selection remains a major weakness: InterCRS reaches a high $CGE^2$ of 0.570, indicating difficulty in deciding when to ask questions rather than recommend items. Structured representations alone are insufficient, as RA-CRS still reports high $FGE^2$ and $CGE^2$ values of 0.333 and 0.461 despite using JSON-format contexts. Likewise, MCTS-based methods do not guarantee reliable planning, with SAPIENT-LLM and T-EPL obtaining $CGE^2$ values of 0.443 and 0.469. Fine-grained elicitation is also unreliable, as reflected by InterCRS's $FGE^2$ of 0.360. For preference exploitation, InterCRS further shows a high $PE^2$ of 0.240, suggesting that explicitly stated preferences are not effectively converted into retrieval cues. Overall, these results indicate that neither JSON-style representation nor generic MCTS planning is sufficient; CRS requires context modeling that jointly supports state tracking, action selection, and preference-grounded retrieval.

\begin{figure*}
    \centering
    \includegraphics[width=1.0\linewidth]{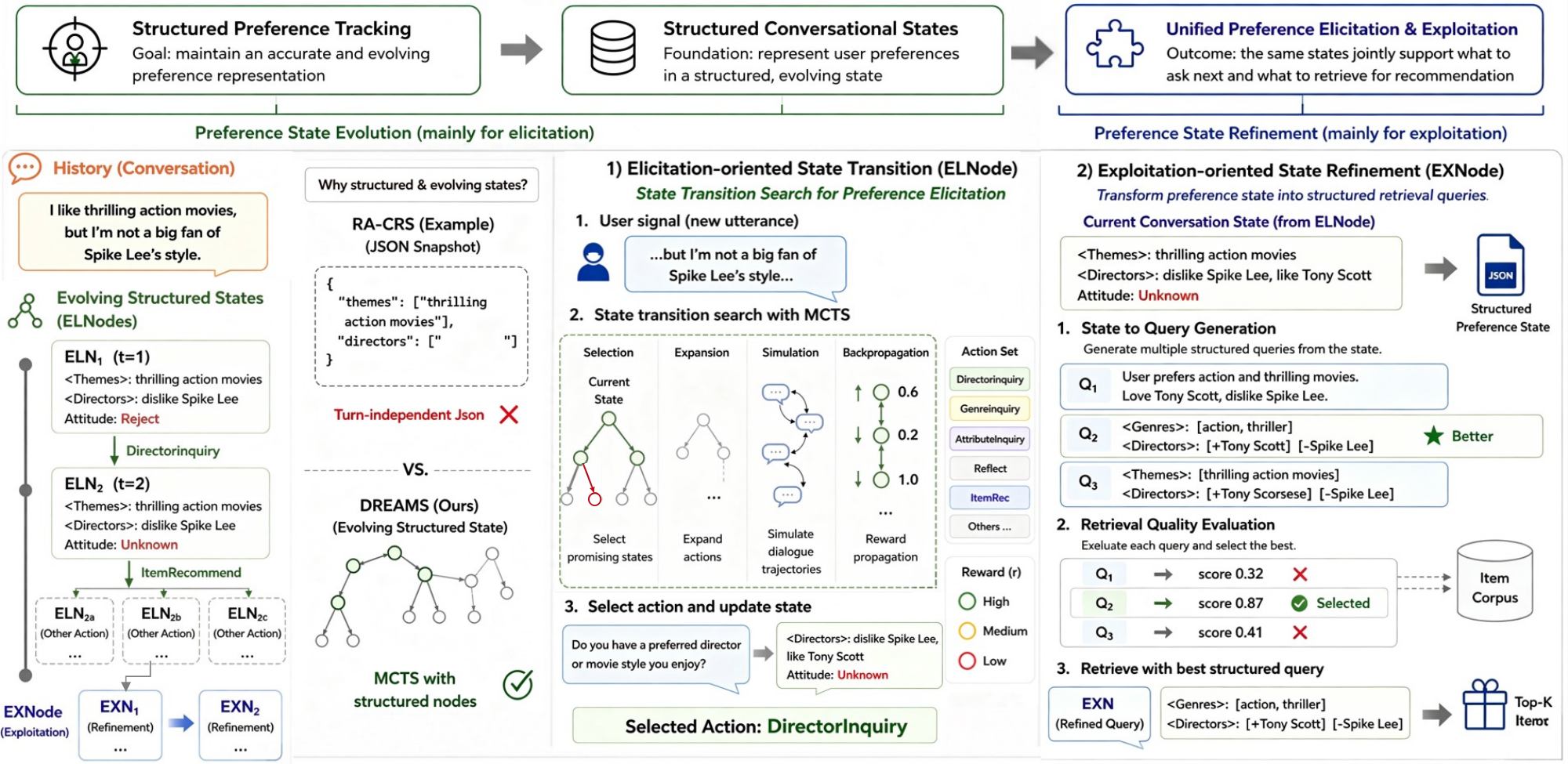}
    \caption{Overview of \ours's structured modeling and unified elicitation \& exploitation. It incorporates ELNodes to explicitly capture estimated user preferences, and EXNodes to enable structured query refinement. }
    \label{fig:method}
\end{figure*}

\section{\ours}
Following prior work \cite{du-etal-2025-sapient, yu2023promptbasedmontecarlotreesearch}, we formulate the CRS interaction as a Markov Decision Process (MDP) $\langle \mathcal{S}, \mathcal{A}, \mathcal{R}, \mathcal{P}, \gamma \rangle$, where the state space $\mathcal{S}$ represents dialogue context and preference information, and the action space $\mathcal{A}$ contains system actions such as asking clarification questions, reflecting on failed recommendations, and recommending items.

\subsection{Overview of \ours}
\noindent\textbf{\ours~Formalization}. 
As shown in Figure~\ref{fig:method}, \ours\ formulates user preference tracking as search over structured conversational states, where nodes encode evolving preference states and edges represent state transitions. 
Unlike JSON-style trackers\footnote{This JSON-style trackers are widely used to make dialogue context explicit and machine-readable \cite{trukhina2026compress, wu2023semantic}.}, which mainly use structured states as turn-independent static memory, \ours\ actively searches over these states for decision-making. It also differs from MCTS-based CRSs such as SAPIENT and T-EPL, whose search nodes are not structured and reusable preference states for open-ended dialogue.
Here, \emph{elicitation} and \emph{exploitation} denote two functions of a CRS rather than the exploration--exploitation trade-off in standard reinforcement learning. Different from prior work that mainly focuses on either elicitation or exploitation, \ours\ contains two node types to jointly optimizing both preference elicitation and exploitation. Given the latest user utterance and the current structured preference state, an ELNode updates the state and searches over actions that determine whether and what the system should ask, whether it should reflect on a failed recommendation, or whether it should proceed to recommendation. When \texttt{ItemRec} is selected, the corresponding EXNode consumes the same evolving preference state and converts it into retrieval-oriented query representations. The recommendation outcome and subsequent user feedback are then incorporated into the next state update. In particular, a rejection can reactivate an ELNode to correct or further elicit the user preferences before another recommendation is made. Thus, ELNodes and EXNodes are not independent modules; they factorize a unified online policy that jointly determines when and what to ask and how to recommend.

Algorithm~\ref{alg:dreams_core} summarizes the full procedure. We further design a hierarchical and verifiable reward to jointly capture elicitation progress and recommendation success. Detailed formulations are provided in Appendix~\ref{app:reward}.

\subsection{ELNode for Preference Elicitation}
\label{elnode}
ELNode supports preference elicitation by searching over possible next actions before the system responds. Its action space covers three purposes: preference elicitation, failure reflection, and transition to exploitation\footnote{Detailed aciton space in Appendix \ref{app:imple}}. 
Each ELNode maintains a JSON-compatible structured preference state that makes positive, negative, missing, and corrected preferences explicit and machine-readable. An LLM parses free-form user utterances into structured key-value updates; for example, ``I am not a big fan of Spike Lee's style'' is mapped to \texttt{<director>: dislike Spike Lee}. The updated state is then used as the basis for MCTS. Here, the core operations — \textit{Selection, Expansion, Simulation, and Backpropagation} — are implemented as follows.

\noindent\textbf{Selection}. Given state $s_i$, \ours~employs the \textit{Upper Confidence bounds applied to Trees} algorithm\footnote{Known for balancing the exploitation (choosing high-value nodes)
and exploration (choosing less-visited nodes).} \cite{kocsis2006bandit} to select the best action (branch) $a^*$, where $A(s_i)$ is the set of child nodes of $s_i$), $R(s_i, a_i')$ is the total reward accumulated from simulations that passed through all the child nodes of $s_i$ after taking the action $a_i'$, $N(s_i)$ is the visit count for node $s_i$, $f(s_i, a_i)$ is the function that returns the child node of $s_i$ after applying action $a_i$, and $c$ is an exploration constant.
\begin{equation}
\small
    \max\limits_{a_i' \in A(s_i)} \left( \frac{R(s_i, a_i')}{N(f(s_i, a_i'))} + c \cdot \sqrt{\frac{2\log N(s_i)}{N(f(s_i, a_i'))}} \right)
\end{equation}

\noindent\textbf{Expansion}. 
Upon reaching a leaf ELNode, the tree expands by attaching new child nodes representing candidate elicitation actions. To efficiently prune the search space and prioritize promising actions, we leverage an LLM to generate prior probabilities $\mathcal{M}$ over the available actions. This LLM guidance ensures that only actions with a positive predicted score are expanded. To reduce the variance of a single LLM call, we sample the LLM $m$ times and aggregate the resulting priors.

\noindent\textbf{Simulation}.
To estimate the long-term effectiveness of an expanded action, \ours~rolls out the dialogue based on a probability-guided policy provided by the LLM until a terminal state or a predefined depth limit is reached. The simulator is initialized with the preference state stored in the current ELNode, and subsequent actions are sampled from the LLM-guided policy. Each transition receives the hierarchical reward described in Appendix~\ref{app:reward}, discounted by $\gamma$.

\noindent\textbf{Back-propagation}.
After each rollout, \ours\ back-propagates the discounted cumulative reward to all nodes on the simulated path and updates their visit counts and expected rewards:
\begin{equation}
\small
    R(s_i, a_i) = \frac{1}{K}\sum\limits_{j=1}^{K} \left( \sum\limits_{s' \in S^j, a' \in A^j} \gamma \cdot r(s', a')\right)
\end{equation}
The reward $r(\cdot)$ consists of three parts: an attitude reward for successful recommendations, an information acquisition reward for learning new preference details, and a turn penalty for conversation efficiency\footnote{See Appendix \ref{app:reward} for details}. After a fixed number of MCTS iterations, \ours~selects the final action by combining the tree value with the LLM prior:
\begin{equation}
    \small
    \max\limits_{a} \left(W_{tree} \cdot \frac{N(f(s_0, a))}{\sum N(f(s_0))} + W_{LLM} \cdot \mathcal{M}(s_0, a)\right)
\end{equation}
where $s_0$ is the root state, and $W_{\mathrm{tree}}$ and $W_{\mathrm{LLM}}$ control the contributions of search statistics and LLM priors.

\subsection{EXNode for Preference Exploitation}
\label{exnode}
When \texttt{ItemRec} isselected, it expands an EXNode to convert the tracked preference state into a retrieval-friendly query with following two steps:

\begin{table*}[t]
    \centering
    \resizebox{0.98\textwidth}{!}{
    \begin{tabular}{l|cccc|cccc|cccc}
        \toprule
        \multirow{2}{*}{Model} 
        & \multicolumn{4}{c|}{Redial (Movie)} 
        & \multicolumn{4}{c}{OpendialKG (Movie)} & \multicolumn{4}{c}{OpendialKG (Book)} \\
        & R@1($\uparrow$) & R@5($\uparrow$) & R@10($\uparrow$) & SR($\uparrow$) 
        & R@1($\uparrow$) & R@5($\uparrow$) & R@10($\uparrow$) & SR($\uparrow$) & R@1($\uparrow$) & R@5($\uparrow$) & R@10($\uparrow$) & SR($\uparrow$) \\
        \midrule
        \multicolumn{13}{c}{\textit{PLM-based}} \\
        \midrule
        BARCOR \cite{wang2022barcor} & 0.193 & 0.467 & 0.593 & 0.107 & 0.094 & 0.383 & 0.494 & 0.211 & 0.106 & 0.375 & 0.456 & 0.219 \\
        UniCRS \cite{wang2022towards} & 0.147 & 0.373 & 0.520 & 0.133 & 0.233 & 0.428 & 0.583 & 0.250 & 0.225 & 0.406 & 0.563 & 0.231 \\
        \midrule
        \multicolumn{13}{c}{\textit{LLM-based}} \\
        \midrule
        InterCRS \cite{wang2023rethinking}& 0.213 & 0.370 & 0.500 & 0.313 & 0.439 & 0.650 & 0.783 & 0.494 & 0.356 & 0.469 & 0.594 & 0.394 \\
        MACRS \cite{fang2024multi}& 0.310 & 0.410 & 0.560 & 0.400 & 0.506 & \underline{0.672} & \underline{0.800} & 0.522 & 0.506 & \underline{0.638} & 0.681 & 0.544 \\
        ChatCRS \cite{li2024incorporating}& \underline{0.367} & \underline{0.473} & \underline{0.573} & \underline{0.440} 
                & 0.539 & 0.650 & 0.772 & 0.550 & 0.494 & 0.600 & 0.675 & 0.531 \\
        PC-CRS \cite{qin2024beyond}& 0.307 & 0.427 & 0.493 & 0.407 & \underline{0.556} & 0.656 & 0.772 & \underline{0.567} & \underline{0.519} & 0.625 & \underline{0.694} & \underline{0.563} \\ 
        RA-CRS \cite{kemper2024retrieval}& 0.273 & 0.340 & 0.387 & 0.333 
               & 0.422 & 0.606 & 0.733 & 0.433 & 0.494 & 0.575 & 0.644 & 0.519\\ 
        SAPIENT-LLM \cite{du-etal-2025-sapient}& 0.287 & 0.387 & 0.427 & 0.380 
               & 0.433 & 0.644 & 0.661 & 0.450 & 0.488 & 0.538 & 0.585 & 0.506\\
        T-EPL \cite{dao-etal-2024-experience}& 0.293 & 0.367 & 0.387 & 0.327 
               & 0.416 & 0.611 & 0.639 & 0.427 & 0.475 & 0.550 & 0.619 & 0.494\\ \midrule
        \ours~(\textit{Ours}) & \textbf{0.507} & \textbf{0.633} & \textbf{0.747} & \textbf{0.560} 
                    & \textbf{0.600} & \textbf{0.728} & \textbf{0.844} & \textbf{0.639} & \textbf{0.550} & \textbf{0.656} & \textbf{0.719} & \textbf{0.594} \\
        
        \bottomrule
    \end{tabular}
    }
    \caption{Overall performance of different methods.}
    \label{tab:main_exp}
\end{table*}

\noindent\textbf{Refinement Mechanism.} 
The EXNode takes the current dialogue history, the JSON-style preference state, and the raw user query as input. It first summarizes the cumulative preference state $s_t^R$, which differs from the ELNode state $s_t$: $s_t$ focuses on the latest turn-level update, while $s_t^R$ aggregates user preferences across the whole dialogue.
Given $s_t^R$, the LLM proposes refinement actions $a_i^R$, such as removing redundant context, reordering preference attributes, and converting informal expressions into machine-readable constraints. For example, a negative preference such as ``dislike Spike Lee's style'' can be converted into \texttt{<director>: dislike Spike Lee}. Applying the $i$-th refinement action produces a refined query state:
\begin{equation}
    EXN_1^i \leftarrow a_i^R(EXN_0)
\end{equation}

\noindent\textbf{Evaluation and Selection.}
DREAMS then retrieves items with each refined query and scores the result by matching the top retrieved items against the preferred attributes in $s_t^R$:
\begin{equation}
\label{eq5}
    R^R(EXN_1^i) = \text{score}(\text{retrieve}(EXN_1^i))
\end{equation}
The query with the highest retrieval score is used for final recommendation generation. In this way, EXNode connects preference tracking to preference exploitation: the system doesn't retrieve from noisy dialogue history directly, but from a structured query derived from evolving preference state.

\section{Experiments}
\label{expr}

\subsection{Experimental Setup}\label{main_settings}

\noindent\textbf{Baselines}. Beyond LLM-based CRSs in Section \ref{evaluation_protocol}, we follow previous studies \cite{qin2024beyond, wang2023rethinking} and involve two pre-trained language model based CRSs (BARCOR \cite{wang2022barcor} and UniCRS \cite{wang2022towards}) and two MCTS-based CRS baselines (SAPIENT \cite{du-etal-2025-sapient} and T-EPL \cite{dao-etal-2024-experience}). See Appendix \ref{app_baselines} for details.

\noindent\textbf{Datasets \& User Simulator}. In addition to the datasets and user simulators in Section \ref{evaluation_protocol}, we follow current common practice in the field \cite{qin2024beyond, huang2024concept, chen2025evaluating} and incorporate OpendialKG \cite{moon2019opendialkg}, a multi-domain benchmark dataset.

\noindent\textbf{Evaluation Metrics.} In addition to the metrics in Section \ref{evaluation_protocol}, we incorporate a human evaluation, which leverages human judgment to further assess the performance of CRSs, specifically focusing on preference elicitation and exploitation. 

\subsection{Main Results}
\label{main_results}

\begin{figure}[t]
    \centering
    \includegraphics[width=0.85\linewidth]{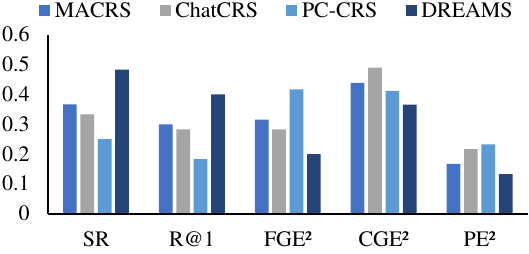}
    \caption{Results on human interaction and evaluation.}
    \label{fig:humaneval}
\end{figure}

\noindent\textbf{\ours~consistently outperforms existing CRS baselines.} Table~\ref{tab:main_exp} shows that \ours~achieves the best performance across all datasets and metrics. Relative to ChatCRS, the strongest non-MCTS baseline, \ours~improves average R@1 and SR by 8.57\% and 9.07\%, respectively. Furthermore, it substantially surpasses the MCTS-based methods SAPIENT-LLM and T-EPL in SR. This suggests that our improvement stems not merely from search, but from grounding search in structured preference states and retrieval refinement.

\begin{table*}[t]
    \centering
    \resizebox{0.98\textwidth}{!}{
    \begin{tabular}{l|cccc|cccc|cccc}
        \toprule
        \multirow{2}{*}{Model} & \multicolumn{4}{c}{Redial (Movie)} & \multicolumn{4}{c}{OpendialKG (Movie)} & \multicolumn{4}{c}{OpendialKG (Book)}\\
        & SR$\uparrow$ & \se$\downarrow$ & \ratio$\downarrow$ & \re$\downarrow$ 
        & SR$\uparrow$ & \se$\downarrow$ & \ratio$\downarrow$ & \re$\downarrow$
        & SR$\uparrow$ & \se$\downarrow$ & \ratio$\downarrow$ & \re$\downarrow$
        \\
        \midrule
        MACRS \cite{fang2024multi}
        & 0.400
        & 0.187 
        & 0.456
        & 0.220 
        & 0.522
        & 0.150
        & 0.431
        & \underline{0.106}
        & \underline{0.544}
        & \underline{0.156}
        & 0.498
        & \underline{0.088}
        \\
        ChatCRS \cite{li2024incorporating}
        & \underline{0.440} 
        & \underline{0.120}
        & 0.561
        & \underline{0.200} 
        & \underline{0.550}
        & \underline{0.133}
        & \textbf{0.291}
        & 0.117
        & 0.531
        & 0.169
        & \underline{0.455}
        & 0.100 \\
        RA-CRS \cite{kemper2024retrieval}
        & 0.333 
        & 0.333
        & 0.461
        & 0.240 
        & 0.433 
        & 0.161
        & 0.398
        & 0.122 
        & 0.519
        & 0.188
        & 0.469 
        & 0.106 \\
        SAPIENT-LLM \cite{du-etal-2025-sapient}
        & 0.380 
        & 0.307
        & \underline{0.443}
        & 0.233 
        & 0.450 
        & 0.150
        & 0.348
        & 0.139 
        & 0.506
        & 0.175
        & 0.458
        & 0.119 \\
        T-EPL \cite{dao-etal-2024-experience}
        & 0.327 
        & 0.313
        & 0.469
        & 0.253 
        & 0.427 
        & 0.144
        & 0.332
        & 0.133
        & 0.494
        & 0.194
        & 0.471 
        & 0.131 \\ \hline
        \ours
        & \textbf{0.560}
        & \textbf{0.100}
        & \textbf{0.289}
        & \textbf{0.160} 
        & \textbf{0.639}
        & \textbf{0.117}
        & 0.304
        & \textbf{0.078}
        & \textbf{0.594}
        & \textbf{0.138}
        & \textbf{0.408}
        & \textbf{0.069} \\
        \bottomrule
    \end{tabular}
    }
    \caption{Elicitation and exploitation evaluation of \ours~and leading baselines. }
    \label{tab:main_exp_ep}
\end{table*}

\noindent\textbf{\ours~exhibits superior performance in preference state tracking, demonstrating enhanced both elicitation and exploitation of user preferences.} As shown in Table~\ref{tab:main_exp_ep}, \ours~consistently yields lower elicitation and exploitation errors than all baselines, reflected by its superior \ratio, \se, and \re~scores. The lower \ratio~indicates fewer coarse-grained decision errors in whether to continue elicitation or proceed to recommendation, while the lower \se~demonstrates stronger fine-grained preference modeling. These gains further lead to lower exploitation errors (\re), suggesting that the inferred preference state can be more effectively transformed into retrieval queries. 
A more specific conclusion is that \ours~ does not benefit from MCTS in isolation. Instead, its gains come from using tree search over a structured preference state, so the system can decide when to clarify, when to recommend, and how to refine the retrieval query from the tracked preferences.

\noindent\textbf{\ours~enjoys enhanced practical utility in real-world human interactions}. 
To evaluate practical utility, we conduct a human study involving 60 participants, who interact with \ours~and three representative baselines (MACRS, ChatCRS, and PC-CRS) under predefined preference settings. As shown in Figure~\ref{fig:humaneval}, \ours~consistently achieves the best performance across all metrics, including SR, Recall@1, \ratio, \se, and \re, indicating superior preference elicitation, preference exploitation, and recommendation effectiveness in realistic conversational settings. To validate the reliability of our automatic evaluator, we randomly sample 50 dialogues generated by \ours~on Redial and ask 10 human annotators to label \re, \se, and \ratio~following the same criteria. The resulting Krippendorff's $\alpha$ scores are 0.74 and 0.68 for Elicitation Error and Exploitation Error, respectively, indicating substantial agreement and supporting the reliability of the automatic evaluation, consistent with prior findings~\cite{liu2023g}.

\noindent\textbf{\ours~generalizes across LLM backbones.}
Table~\ref{tab:gemini} shows that \ours~maintains clear advantages on both Gemini-2.5-Flash and GPT-4o-mini. Despite Gemini's strong long-context capability, the free-form method InterCRS still suffers from high elicitation errors, whereas \ours~consistently achieves better preference tracking and recommendation performance. These results suggest that the limitation of free-form context modeling is not primarily due to context length. Instead, effective conversational recommendation requires an explicit preference state that supports preference elicitation and exploitation.

\subsection{Recommendation Performance Analysis}\label{in-depth}

\begin{table*}[ht]
    \centering
    \resizebox{0.98\linewidth}{!}{
    \begin{tabular}{l|ccccc|ccccc|ccccc}
        \toprule
        \multirow{2}{*}{Model} 
        & \multicolumn{5}{c|}{Redial (Movie)} 
        & \multicolumn{5}{c}{OpendialKG (Movie)}
        & \multicolumn{5}{c}{OpendialKG (Book)}\\
        & R@1($\uparrow$) & SR($\uparrow$) & \se~($\downarrow$) & \ratio($\downarrow$) & \re~($\downarrow$)
        & R@1($\uparrow$) & SR($\uparrow$) & \se~($\downarrow$) & \ratio($\downarrow$) & \re~($\downarrow$) 
        & R@1($\uparrow$) & SR($\uparrow$) & \se~($\downarrow$) & \ratio($\downarrow$) & \re~($\downarrow$)\\
        \midrule
        \ours & \textbf{0.507} & \textbf{0.560} & \textbf{0.100} & \textbf{0.289} & \textbf{0.160}
                    & \textbf{0.600} & \textbf{0.639} & \textbf{0.117} & \textbf{0.304} & \textbf{0.078} &
                    \textbf{0.550} &
                    \textbf{0.594} &
                    \textbf{0.138} &
                    \textbf{0.408} &
                    0.069\\
        \midrule
        \multicolumn{16}{c}{\textit{Structured Context Modeling}} \\
        \midrule
        w/o EXNode & 0.407 & 0.487 & 0.133 & 0.311 & 0.193 & 0.517 & 0.572 & 0.139 & 0.326 & 0.117 & 0.494 & 0.556 & 0.144 & 0.422 & 0.100 \\
        w/o ELNode & 0.393 & 0.427 & 0.173 & 0.417 & 0.173 & 0.500 & 0.550 & 0.156 & 0.378 & 0.094 & 0.444 & 0.456 & 0.188 & 0.490 & \textbf{0.063} \\
        w/o ELNode \& Json & 0.287 & 0.347 & 0.340 & 0.539 & 0.233 & 0.450 & 0.478 & 0.406 & 0.455 & 0.111 & 0.388 & 0.456 & 0.250 & 0.523 & 0.075 \\
        \midrule
        \multicolumn{16}{c}{\textit{Structured Tracking}} \\
        \midrule
        w/ Json Only & 0.327 & 0.360 & 0.200 & 0.455 & 0.240 & 0.439 & 0.472 & 0.411 & 0.384 & 0.128 & 0.394 & 0.475 & 0.213 & 0.473 & 0.081 \\
        \bottomrule
    \end{tabular}
    }
    \caption{Ablation studies on benchmark datasets.}
    \label{tab:ablation}
\end{table*}
We analyze the contribution of different components through the following variants:
\begin{itemize}[leftmargin=*]
    \item \underline{w/o EXNode}. Directly retrieves from free-form dialogue history without query refinement.
    \item \underline{w/o ELNode}. Replaces MCTS-based state transition with direct LLM action selection.
    \item \underline{w/o ELNode \& Json}. Further removes structured preference states and tracks preferences using free-form context.
    \item \underline{w/ Json Only.} Uses JSON-style preference states, but performs action selection and retrieval without structured search.
\end{itemize}

Table \ref{tab:ablation} disentangles the effects of \textit{structured preference states}, \textit{state-transition search}, and \textit{structured retrieval refinement}. Removing both ELNode and JSON causes the largest degradation, significantly increasing \se and \ratio, showing that explicit structured states are essential for effective preference tracking. Removing ELNode while preserving JSON also degrades performance, indicating that preference states must be actively updated and utilized during interaction. Replacing MCTS with direct LLM reasoning further reduces SR and increases \se, suggesting that structured states should serve as actionable decision states rather than passive memory. Removing EXNode mainly decreases Recall@1 and SR while increasing \re, demonstrating the importance of exploitation-side query refinement. Overall, the results show that structured preference states, state-transition search, and structured retrieval refinement are all critical to unified preference elicitation and exploitation.

\subsection{Preference Elicitation Analysis}
\noindent\textbf{Structured conversational states improve preference elicitation by transforming noisy dialogue history into stable and decision-relevant preference representations}. 
Table \ref{tab:ablation} supports this conclusion through the ELNode and JSON-related ablations. Removing ELNode weakens the model's ability to maintain evolving preference states, increasing both \se and \ratio. Further removing JSON-style states leads to even larger degradation. These results show that explicit structured state modeling is essential for effective preference elicitation, enabling \ours~to outperform MCTS-based baselines such as SAPIENT-LLM and T-EPL. Case study (See Appendix \ref{app:case_study}) illustrates how \ours~differs from baselines facing noisy dialogue history.

\noindent\textbf{JSON-format state tracking records preferences explicitly, but structured search is needed to turn the recorded state into better decision making}. 
Both RA-CRS and the w/Json Only variant maintain structured preference states, yet neither operationalizes them through structured search. Table \ref{tab:ablation} isolates this effect: although the w/Json Only variant preserves the same structured states and action/query interfaces, replacing MCTS with direct LLM reasoning substantially lowers SR and increases \se. This demonstrates that effective preference tracking requires not only structured states, but also search over evolving conversational states for lookahead decision-making. A mechanism-level case study is provided in Appendix \ref{app:case_study}.

\subsection{Preference Exploitation Analysis}

\begin{figure}[t]
    \centering
    \includegraphics[width=0.45\textwidth]{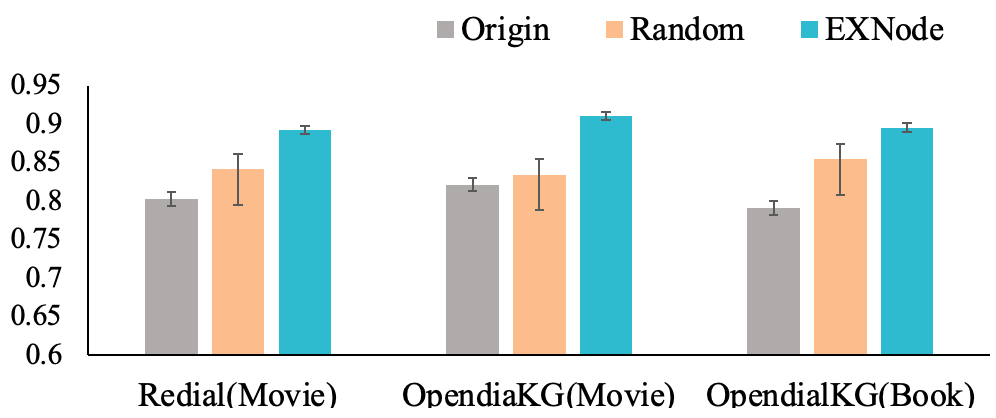}
    \caption{Comparison of query refinement methods via cosine similarity to ground-truth embeddings.}
    \label{fig:embeddings}
\end{figure}

\noindent\textbf{By leveraging the refinement mechanism, the EXNode progressively refines the retrieval queries, steering them closer to the ground-truth item description}.  
As shown in Figure \ref{fig:embeddings}, EXNode progressively refines retrieval queries toward the ground-truth item description, yielding consistently higher cosine similarity than original queries. In contrast, randomly paraphrased LLM queries are unstable and can even underperform the original query, highlighting the sensitivity of retrieval to contextual noise \cite{huang2024concept}. We further observe that EXNode tends to generate more structured JSON-style retrieval queries, which better align with the structured knowledge bases in REDIAL and OpenDialKG. Therefore, EXNode improves exploitation through both semantic refinement and structured retrieval representations.

\subsection{Efficiency Analysis}
\label{efficiency}
Although a common concern with tree-based planning is latency, Table \ref{tab:dreams_hyperparameter} shows that vanilla \ours~already achieves efficiency comparable to existing CRSs through our parallel function calling adaptation. To further improve deployment efficiency, we introduce \textbf{\ours(EA)}, an \underline{E}xperience-\underline{A}ugmented variant inspired by prior experience-based reasoning methods \citep{gu-etal-2025-toward, ren2025sigmarefininglargelanguage}. Specifically, we store MCTS trajectories as structured experience knowledge and retrieve structurally similar successful/failed cases from an Experience Knowledge Base (EKB) to warm-start decision making, thereby bypassing expensive online tree search. As shown in Table \ref{tab:dreams_hyperparameter}, \ours(EA) reduces inference latency to 9s while maintaining competitive recommendation performance (Recall@1=0.467). More implementation and efficiency details are provided in Appendix~\ref{app:efficiency}.

\section{Conclusion}
In this paper, we highlight the critical role of structured context modeling in conversational recommender systems and propose \ours, a dual-node MCTS framework that jointly optimizes preference elicitation and preference exploitation. Extensive experiments demonstrate that effective conversational recommendation requires not only structured representations, but also structured search over evolving conversational states. We hope this work encourages future CRS research to treat structured context as an important active decision mechanism.

\section*{Limitations}
To ensure a fair comparison, we adhered to established protocols \cite{qin2024beyond, li2024incorporating} by limiting all methods to the GPT-4o-mini and Gemini-2.5-flash as their backbone. Due to constraints in computational resources and budget, we do not extend our evaluation to other SOTA LLMs (e.g., Claude Opus). Additionally,
\ours~ may inherit model-level limitations of its LLM backbone, including social biases and incomplete domain knowledge. The scope of this work is not to remove these intrinsic limitations, but to improve preference tracking and recommendation through a model-agnostic structured-state framework. We encourage future research to investigate the influence of diverse LLM backbones on CRS performance.

More broadly, the evaluation of modern CRSs remains challenging because scalable protocols often depend on LLM-based judges, either as offline evaluators or as user simulators. Although such protocols enable controlled and reproducible comparisons, they may inherit biases from the evaluator model, prompt design, and predefined user preference ontology. Finally, generalization to domains with different item structures, knowledge requirements, or decision costs remains untested. Future work should develop more reliable and diverse human-grounded benchmarks, behaviorally validated user simulators, and evaluator-independent protocols for conversational recommendation.

\section*{Acknowledgments}
This research is supported by the National Research Foundation, Singapore under its National Large Language Models Funding Initiative (AISG Award No: AISG-NMLP-2024-002), the Singapore Ministry of Education (MOE) Academic Research Fund (AcRF) Tier 1 grant (Proposal ID: 24-SIS-SMU-002), and National Natural Science Foundation of China (No. 62272330 and No.U24A20328). Yang Deng is supported by the Lee Kong Chian Fellowship awarded by Singapore Management University. Any opinions, findings and conclusions or recommendations expressed in this material are those of the author(s) and do not reflect the views of National Research Foundation, Singapore.

\bibliography{anthology,custom}

\appendix
\section{Implementation Details}
\label{app:imple}

All experiments are conducted using a single Nvidia RTX A6000 GPU. 

\subsection{Implementation of baselines}
\label{app_baselines}
Baseline implementations are sourced from their respective official code repositories on GitHub or publicly available checkpoints. GPT-4o-mini is employed as the LLM backbone and \textit{text-embedding-ada-002} is used for recommendation module for all the LLM-based CRSs.

Notably, We adapt the two MCTS-based CRS baselines(SAPIENT\cite{du-etal-2025-sapient} and T-EPL\cite{dao-etal-2024-experience}) to our dialogue-based LLM-powered setting through a unified high-level planning interface. 

\begin{itemize}[leftmargin=*]
    \item \textbf{SAPIENT-LLM.} Consider that SAPIENT is originally designed for attribute-based CRS with discrete ask/recommend actions over attribute values and items. To adapt it to our dialogue-based LLM-powered CRS setting, we replace the original attribute-value state with an LLM-maintained dialogue state, including the conversation history, extracted positive and negative user preferences, rejected items, and the current candidate item set. We preserve its high-level action space, where ask selects a preference facet inferred from item metadata and dialogue context, and recommend selects items from the current candidate ranking. The selected abstract action is then verbalized by the LLM into a natural-language response, while the subsequent user utterance is parsed by the LLM state tracker to update the dialogue state.
    \item \textbf{T-EPL.} Consider that T-EPL is originally formulated for target-driven recommendation dialogues and assumes access to a predefined target item. To make it comparable in our dialogue-based CRS setting, we remove access to the ground-truth target item and replace its target-conditioned value function with a belief-weighted value over the current candidate distribution. Specifically, the base recommender produces a ranked candidate set given the current dialogue state, and T-EPL estimates the value of a simulated state by retrieving similar past dialogue states from memory and aggregating their outcome scores weighted by the current candidate probabilities. During MCTS expansion, candidate actions are represented as abstract dialogue acts, such as asking about an uncertain preference, recommending an item, clarifying ambiguous constraints, or justifying a recommendation, and are finally realized as natural-language responses by the LLM.
\end{itemize}

\subsection{Difference between \ours~and baselines}
\label{app:comparison}
As shown in Table \ref{tab:comparison}, compared with each baseline, \ours~ differs as follows: \textbf{SAPIENT} uses MCTS with free-form nodes only for attribute elicitation and relies on free-form exploitation; \textbf{T-EPL} similarly restricts free-form-node MCTS to target-oriented topic planning; \textbf{InterCRS} relies entirely on free-form dialogue to evaluate LLM performance; \textbf{MACRS} focuses on agent collaboration in free-form dialogue; \textbf{PC-CRS} studies persuasion and credibility under free-form interaction; \textbf{ChatCRS} enhances domain knowledge while retaining free-form dialogue; and \textbf{RA-CRS} uses turn-independent JSON only for preference elicitation, followed by free-form exploitation. In contrast, \ours~ jointly models preference elicitation and exploitation using MCTS over structured conversational states.

\begin{table*}[t]
\centering
\small
\setlength{\abovecaptionskip}{5pt}
\setlength{\belowcaptionskip}{0pt}
\begin{tabularx}{\linewidth}{l p{3.5cm} p{3.5cm} X p{1.1cm}} 
\toprule
\textbf{Model} & \textbf{Elicitation} & \textbf{Exploitation} & \textbf{Design Highlights} & \textbf{Task} \\ 
\midrule 

SAPIENT 
& MCTS with free-form node 
& Free-form (whole dialogue) 
& Free-form node MCTS just for elicitation. 
& Attr. \\

\midrule 

T-EPL 
& MCTS with free-form node 
& Free-form (whole dialogue) 
& Free-form node MCTS just for topic planning.  
& Target \\

\midrule 

InterCRS 
& Free-form (whole dialogue) 
& Free-form (whole dialogue) 
& LLMs' performance evaluation. 
& Dialog \\

MACRS 
& Free-form (whole dialogue) 
& Free-form (whole dialogue) 
& Agents collaboration. 
& Dialog \\

PC-CRS 
& Free-form (whole dialogue) 
& Free-form (whole dialogue) 
& Persuasiveness and credibility study. 
& Dialog \\

ChatCRS 
& Free-form (whole dialogue) 
& Free-form (whole dialogue) 
& Domain-specific knowledge enhancement. 
& Dialog \\

RA-CRS 
& Turn-independent JSON 
& Free-form (whole dialogue) 
& JSON for preference elicitation. 
& Dialog \\

\rowcolor[HTML]{EBF4FF} 
\ours 
& MCTS with structured node 
& MCTS with structured node 
& MCTS for both elicitation and exploitation with structured node 
& Dialog \\ 

\bottomrule
\end{tabularx}

\caption{
Comparison of conversational recommendation methods from the perspective of preference elicitation and exploitation. Existing approaches primarily rely on free-form dialogue context or partially structured elicitation, whereas \ours jointly models elicitation and exploitation through MCTS over structured conversational states.
}
\label{tab:comparison}
\end{table*}

\subsection{Implementation of \ours}\label{app:ours}
In this section, we will introduce in detail the implementation of different components in \ours~, including: actions for ELNode transition, state update component, LLM scorer, online simulator, reward function and EXNode refinement.

\subsubsection{Actions for ELNode transition}
\ours~ employs an iterative Monte Carlo search algorithm to identify the optimal action at each round from the action set defined in Table~\ref{tab:action_set}. 
{The action space is organized around the core policy decisions in multi-turn conversational recommendation: acquiring missing preferences, resolving ambiguity or recovering from negative feedback, recommending from the current preference state, and responding to a request for item details. This design follows the ask--recommend--feedback structure studied in prior CRS policies. EAR integrates preference estimation, system action, and reflection over user feedback \cite{lei2020estimation}, while UNICORN learns a unified policy over asking about attributes and recommending items \cite{deng2021unified}. \ours~ instantiates these decisions at a finer granularity for LLM-based CRS: attribute-specific inquiry actions acquire or disambiguate preferences, \textsc{FailureReflection} handles rejected recommendations, \textsc{ItemRecommendation} activates EXNode-based retrieval, and \textsc{ItemExplanation} handles post-recommendation information requests. Table~\ref{tab:action_space} summarizes the role and state effect of each action.}

The specific implementation prompts associated with each action are provided in Appendix~\ref{prompts}.

\begin{table*}[t]
\centering
\small
\begin{tabular}{p{0.17\linewidth}p{0.23\linewidth}p{0.50\linewidth}}
\toprule
Action & Policy role & Trigger and effect \\
\midrule
\textsc{GenreInquiry} & Preference acquisition & Asks for a missing or ambiguous genre preference and updates the corresponding state field. \\
\textsc{StarInquiry} & Preference acquisition & Asks for a missing or ambiguous actor preference and updates the corresponding state field. \\
\textsc{DirectorInquiry} & Preference acquisition & Asks for a missing or ambiguous director preference and updates the corresponding state field. \\
\textsc{FailureReflection} & Feedback recovery & Uses a rejected recommendation and its stated reason to correct the preference state and select a suitable follow-up action. \\
\textsc{ItemRec} & Transition to exploitation & Activates an EXNode, which refines the structured state into a retrieval query, retrieves and reranks candidates, and produces a recommendation. \\
\textsc{ItemExplanation} & Post-rec response & Provides requested information about an already recommended item without prematurely initiating a new recommendation. \\
\bottomrule
\end{tabular}
\caption{Action space of the ELNode policy. The actions cover preference acquisition, recovery from negative feedback, transition to preference exploitation, and post-recommendation explanation.}
\label{tab:action_space}
\end{table*}

\subsubsection{State Update Component}
In order to dynamically and adaptively model user preferences, \ours~ uses a large model to structure free-form text to achieve continuous updating of the conversation state.

\subsubsection{LLM Scorer}
The LLM scorer is used for providing prior probabilities over possible actions for the tree node expansion and the optimal action selection after iterations. To enhance robustness, we sample the LLM’s predictions $m$ times and aggregate the results into a probability distribution. See Figure \ref{fig:llm_score} for details and prompts of each prediction time.

\subsubsection{Online Simulator}
During simulation of planning-tree, we conduct an online user simulator to give feedback in response to simulated actions. 

\subsubsection{Pseudo-Code}
We provide the pseudo-code of \ours\ in Algorithm \ref{alg:dreams_core}.

\begin{algorithm}[h!]
\small
\SetAlgoLined
\DontPrintSemicolon
\SetKwInOut{Input}{Input}
\SetKwInOut{Output}{Output}
\SetKwFunction{MCTSSimulate}{MCTS\_Simulate\_and\_Backpropagate}
\SetKwFunction{ELNodeExpand}{ELNode\_Expand}
\SetKwFunction{EXNodeRefineSearch}{EXNode\_Refine\_Search}
\SetKwFunction{StateTracking}{State\_Tracking}
\SetKwFunction{RetrieveItem}{Item\_Retrieve}

\Input{Iteration Steps $n(El)$ and $n(Ex)$, Simulation Depth $k(El)$, Max Turn $T$}
\Output{CRS Response}

Set $ELNode_0=\{\}$, Dialogue History $H=\{\}, t=1$\;
\While{dialogue turn $t<T$}{
    Receive User Utterance $u_t$\;
    Update $ELNode_{t}$ using $ELNode_{t-1}$ and $u_t$\; 

    \tcc{1. ELNode Process Start}
    \For{$i \leftarrow 1$ \KwTo $n(El)$}{
        Perform MCTS with max tree depth being $k(El)$, and obtain the estimated value for each action\;
    }
    Obtain best action $a^*$ based on multiple MCTS trials\;
\tcc{2. EXNode Process Start}
    \If{$a^* = \text{ItemRecommendation}$}{
        Initialize query $EXNode_0 = H$\; 
        LLM generates refinement actions $[a^R_1,... a^R_j, ..., a^R_{n(Ex)}]$\;
        Obtain query scores $[R^R_1,... R^R_j, ..., R^R_{n(Ex)}]$ using Eq.\ref{eq5}\;
        Recommend Item $I$ using query with highest score\;
        
        \If{User Accepts $I$}{
            \textbf{Return} Successful Recommendation\;
        }
    }
    \Else{
        Generate Response $r_t$ aligning with $a^*$\;
    }
    
    $H \leftarrow H \cup \{u_t, r_t\}$, $t = t+1$\;
}
\textbf{Return} Dialogue Termination (Max Turns Reached)\;

\caption{Pseudo-code of \ours}
\label{alg:dreams_core}
\end{algorithm}

\subsection{Reward Design}
\label{app:reward}

The reward in \ours~ is designed as a hierarchical combination of intermediate and final signals, rather than a single sparse success indicator. For a simulated transition $(s_t, a_t, s_{t+1})$, we define the immediate reward as:
\begin{equation}
r_t = \lambda_e r_t^{\text{explore}} + \lambda_r r_t^{\text{retrieve}} + \lambda_u r_t^{\text{attitude}} - \lambda_c r_t^{\text{cost}},
\end{equation}
where $r_t^{\text{explore}}$ measures whether the current action improves the completeness of the structured preference state, $r_t^{\text{retrieve}}$ measures whether the refined query leads to retrieval results that are more consistent with the hidden target profile, $r_t^{\text{attitude}}$ captures the user simulator's feedback after interaction, and $r_t^{\text{cost}}$ penalizes unnecessarily long or repetitive trajectories.

The overall action value is computed as the discounted return:
\begin{equation}
R(s_t,a_t)=\mathbb{E}\left[\sum_{k=t}^{T}\gamma^{k-t} r_k + \lambda_f \cdot \mathrm{Accept}_T\right],
\end{equation}
where $\mathrm{Accept}_T \in \{0,1\}$ indicates whether the simulated dialogue ends with user acceptance.

For exploration, $r_t^{\text{explore}}$ is positive when the action helps clarify previously missing or low-confidence attributes in the structured user state. This encourages the planner to ask informative questions when the current preference state is still incomplete.

For exploitation, $r_t^{\text{retrieve}}$ is defined using verifiable signals from the environment. Concretely, we compare the attributes of the top retrieved item with the ground-truth attribute profile underlying the simulator. Let $\hat{y}_t^{(1)}$ denote the top-1 retrieved item and let $A(\cdot)$ denote its attribute set. We compute retrieval reward from the overlap between $A(\hat{y}_t^{(1)})$ and the target attribute set, so that higher reward is assigned only when retrieval becomes more faithful to the actual user preference profile. This makes the retrieval reward externally checkable, instead of relying solely on free-form LLM scoring.

Finally, $r_t^{\text{attitude}}$ rewards trajectories that lead to explicit user acceptance and penalizes rejection, while $r_t^{\text{cost}}$ discourages overly long dialogues. In this way, the reward design combines dense intermediate supervision with a final outcome objective. Although \ours~ does not optimize the model with reinforcement learning, this design is conceptually aligned with recent verifiable-reward based paradigms, in that the key reward terms are grounded in observable and reproducible signals from the environment.

\subsubsection{R-chain Refinement}
The R-chain first processes the multi-turn dialogue history to extract explicit user preferences regarding genres, actors, directors, and thematic elements. This extraction is performed by prompting an LLM with a structured instruction that emphasizes fine-grained disambiguation of closely related genres (e.g., “thriller” vs. “crime”) and returns a structured JSON of liked/disliked entities with associated confidence scores.

Once preferences are extracted, the system formulates candidate retrieval queries using strategies defined by the LLM. A lightweight Monte Carlo Tree is employed to select the optimal refinement strategy. Specifically, a tree of strategy nodes is constructed, where each node corresponds to a paraphrasing strategy and is evaluated based on the simulated performance of the resulting query. During simulation, a query is generated via the LLM based on the node’s strategy and then evaluated using a two-part reward function: (1) a preference match score that quantifies how well top retrieved items align with the user’s stated preferences, and (2) a potential attitude score that estimates user acceptance likelihood based on cumulative preference overlap and diversity of retrieved genres. The combined reward guides backpropagation in the tree, leading to the selection of the most promising strategy.

The final refined query is then embedded and used to retrieve top-5 candidate items via cosine similarity over precomputed item embeddings. The top-ranked items are cached for recommendation, and metadata including the selected strategy, refined query text, and extracted preferences are returned for transparency and downstream explanation. This structured search and modeling approach ensures not only accurate preference alignment but also adaptive query optimization based on retrieval performance feedback.

\subsection{Implementation of Baselines}
For all baselines, we used the official code with the checkpoints from the original paper’s GitHub repository or the official prompt from the original paper for implementation. We refer to Appendix \ref{prompts} for prompts.
For a fair comparison, all LLM-based CRSs including \ours~ employ the gpt-4o-mini\footnote{gpt-4o-mini-2024-07-18} as backbone and the text-embedding-ada-002\footnote{} encoder as the recommendation module to retrieve items.

\subsection{Implementation of Evaluation Metrics}
\label{app_metric}
To systematically assess the preference elicitation and exploitation capability of LLM-based CRSs, we implemented a multi-step evaluation protocol that includes metrics of Fine-Grained Elicitation Error (\se~), Preference Exploitation Error (\re~) and Course-Grained Elicitation Error (\ratio~). Among them, \se~ and \re~ are dialogue-level metrics, and \ratio~ is a turn-level metric. In the following content, we will elaborate on the design details of metrics. How LLM evaluates these metrics as an evaluator will be explained in \ref{app_evaluator}, and the prompts are in \ref{prompts}.

\begin{table*}[ht]
    \centering
    \renewcommand\arraystretch{1.5}
    \resizebox{0.9\textwidth}{!}{%
    \begin{tabular*}{\textwidth}{p{3cm}|p{5.678cm}|p{6cm}}
        \hline
        \multicolumn{3}{c}{\textbf{CRS Evaluation Metrics}} \\ \hline
        \textbf{Metric} & 
        \textbf{Definition} & 
        \textbf{Measurement}\\ \hline
        \makecell[l]{Course-Grained \\ Elicitation \\ Error (\ratio)}
        & 
        \makecell[l]{Error rate of high-level decision (i.e., \\make recommendation and ask for\\ information) mismatch between \\retrieval quality and action taken.}
        &
        \makecell[l]{We run retrieval to see if the retrieval is \\correct and annotate CRS's action \\of each turn within each conversation. \\we measure the proportion of wrong \\actions across all conversations as the \\metric, where wrong action means the \\CRS: 
        \\• Ask when retrieval is correct, or 
        \\• Recommend when retrieval is wrong. 
        \\Exclude turns with other actions.}
        \\ \hline
        \makecell[l]{Fine-Grained \\Elicitation \\Error (\se)}
        &
        \makecell[l]{Error rate of ineffective follow-up \\questions after the system receives \\specific user feedback about their \\preferences. For example, \\when a user expresses dislike for \\certain directors' styles, the CRS fails\\ to ask the right questions (ask about\\ the director) but take a wrong strategy \\(i.e., ask about movie genre or succe-\\ssive recommendation).}
        & 
        \makecell[l]{For each dialogue, consider all turns th-\\at immediately follow user feedback to \\a recommendation. The fine-grained el-\\icitation error is defined as the proporti-\\on of such turns within the dialogue w-\\here the CRS action mismatches the us-\\er intent (e.g., issuing an irrelevant inq-\\uiry or making a premature recommen-\\dation).}
        \\ \hline
        \makecell[l]{Preference\\ Exploitation\\ Error (\re) }
        &
        \makecell[l]{Failure rate of retrieving incorrect it-\\ems after full preference explicit dis-\\closure. For example, known user \\preference on action, adventure and \\drama movies, but recommending a \\thriller movie.} 
        & 
        \makecell[l]{When user simulator reveals all his pre-\\ference (identified by an advanced extra \\LLM), we count the proportion of inco-\\rrect recommendations.} \\ \hline
    \end{tabular*}
    }
    \caption{Key performance indicators for conversational recommendation systems}
    \label{tab:ep_metrics}
\end{table*}

\subsubsection{Dialogue-level Metrics}
\noindent \textbf{Fine-Grained Elicitation Error.} A "Fine-Grained Elicitation Error" (\se) is defined as a failure by the assistant to adapt its elicitation or exploitation strategy in response to explicit negative feedback from the user. If the assistant ignores user rejection of a previous recommendation (e.g., a movie was disliked due to its horror genre) and subsequently issues a recommendation (e.g., continue to recommend horror movies) or inquiry (e.g., inquiry for stars or directors) that does not reflect learning from that rejection, it constitutes a strategic fault.

\noindent Formally, let $t_i$ denote the $i$-th assistant turn, and $r_j$ the last rejected recommendation with user-stated rejection reason $R_j$. Then, a \se~ occurs if:

\begin{equation}
    \begin{split}
        \text{\se}(t_i) = 
        \begin{cases}
        1 & \text{if } \text{intent}(t_i) , \text{ask}\} \land \\ &  \neg \text{reflects}(t_i, R_j) \\
        0 & \text{otherwise}
        \end{cases}
    \end{split}
\end{equation}

\noindent Here, $reflects(t_i, R_j)$ is a context-aware predicate indicating whether the assistant incorporated the rejection reason $R_j$ into its follow-up action (e.g., asking clarifying questions or avoiding similar content).

\noindent \textbf{Preference Exploitation Error.} 
A "Preference Exploitation Error" (\re) is triggered when the assistant has successfully elicited explicit user preferences across all three key attributes — genre, actor, and director — yet the retrieved item associated with the assistant's recommendation is inappropriate (recall@5 = False).

\noindent Let $P = \left\{ p_g, p_a, p_d\right\}$ represent the user's fully specified preferences for genre, actor, and director. Given a dialog context $C$, a \re is defined as:

\begin{equation}
    \begin{split}
        \text{\re}(t_k) =
        \begin{cases}
        1 & \text{if } \text{pref\_satisfied}(C) =  \text{True} \\ & \land \text{retrieval}(t_k) = \text{False} \\
0 & \text{otherwise}
    \end{cases}
    \end{split}
\end{equation}

\noindent The function $pref_satisfied(C)$ returns true only if the assistant has successfully acquired all three preference dimensions 
$p_g$, $p_a$, $p_d$ from the user at or before turn $t_k$

\noindent \textbf{Evaluation Process.} 
Evaluation is conducted in dialog-level granularity. For each conversation file: 1) A GPT-4o model is provided with the full dialog history and a structured system prompt defining the criteria for \se~ and \re~. 2) The model returns a JSON-encoded response:
\{"\se": true/false, "\re": true/false\}.
The evaluation script aggregates the counts of \se$_{true}$, \se$_{false}$, \re$_{true}$, \re$_{false}$ over all dialogs to compute proportions. Final scores are reported as:

\begin{equation}
    \begin{split}
        \text{\se} = \frac{\sum \text{\se}_{\text{true}}}{\sum \text{\se}_{\text{true}} + \sum \text{\se}_{\text{false}}} \\
        \text{\re} = \frac{\sum \text{\re}_{\text{true}}}{\sum \text{\re}_{\text{true}} + \sum \text{\re}_{\text{false}}}
    \end{split}
\end{equation}

\subsubsection{Turn-level Metric}
For the evaluation of \ratio, we categorizes assistant utterances in the recorded data into predefined classes based on their intent and the retrieval relevance of associated results. The classification schema is composed of four primary categories: Ask-True, Ask-False, Recommend-True, Recommend-False. 
\noindent \textbf{Intent and Relevance Classification Framework.}
Each assistant utterance $u_i$ within a conversation $C = \left\{u_1, u_2, ... u_n\right\}$ is analyzed using an LLM-powered evaluator, prompted with structured instructions and context information. The model classifies each utterance based on:
\begin{equation}
    \begin{split}
    &intent(u_i) \in \left\{\text{Ask, Recommend, Others}\right\} \\ 
    &retrieval(u_i) \in \left\{\text{True, False}\right\}
\end{split}
\end{equation}

Using the above classification, we define the following evaluation counts:


\begin{small}
Let \(a_i = \operatorname{intent}(u_i)\) and
\(r_i = \operatorname{retrieval}(u_i)\).
\begin{align}
\mathrm{AT}
&= \sum_i \mathbb{1}
   [a_i=\text{Ask} \land r_i=\text{True}], \\
\mathrm{AF}
&= \sum_i \mathbb{1}
   [a_i=\text{Ask} \land r_i=\text{False}], \\
\mathrm{RT}
&= \sum_i \mathbb{1}
   [a_i=\text{Recommend} \land r_i=\text{True}], \\
\mathrm{RF}
&= \sum_i \mathbb{1}
   [a_i=\text{Recommend} \land r_i=\text{False}].
\end{align}
\end{small}

To enable direct comparison between dialog behavior classes, we compute the relative frequency of each act category among all labeled utterances that fall under either Ask or Recommend intents. Specifically, for each class $x \in \{AT,AF,RT,RF\}$, we calculate its normalized ratio $r_x$ as follows:

\begin{align}
r_{\text{AT}} &= \frac{\text{AT}}{\text{AT} + \text{AF} + \text{RT} + \text{RF}} \\
r_{\text{AF}} &= \frac{\text{AF}}{\text{AT} + \text{AF} + \text{RT} + \text{RF}} \\
r_{\text{RT}} &= \frac{\text{RT}}{\text{AT} + \text{AF} + \text{RT} + \text{RF}} \\
r_{\text{RF}} &= \frac{\text{RF}}{\text{AT} + \text{AF} + \text{RT} + \text{RF}} 
\end{align}

Here, True and False refer to the relevance of the retrieval result, not to the correctness of the assistant action. Therefore, Ask-True denotes an unnecessary elicitation action when retrieval is already sufficient, and Recommend-False denotes a premature recommendation when retrieval is still incorrect. Under this convention, the coarse-grained elicitation error is defined as

\begin{align}
    \text{\ratio} &= r_{\text{AT}} + r_{\text{RF}}
\end{align}

These ratios ensure that evaluation results reflect the distributional tendencies of the assistant’s strategy, independent of the total number of utterances. This normalization is especially useful for comparing across different models or dialog scenarios with unequal total counts.

\subsubsection{Prefix-level Next Action Evaluation}
\label{app:next_action_quality}
Table \ref{tab:next_action_quality} shows a clear progression from state representation to state-guided decision making and then to structured search. RA-CRS uses a JSON-format dialogue state, but its decision policy only weakly operationalizes that state. \ours~ w/Json Only improves over RA-CRS by heuristically using the structured state, increasing Next-Action Acc. from 0.542 to 0.633 and reducing Premature Rec. Rate from 0.308 to 0.217. Full \ours~ further improves all metrics by replacing heuristic use with structured search, reaching 0.758 Next-Action Acc. and reducing Redundant Q. Rate and Premature Rec. Rate to 0.089 and 0.117. These results indicate that structured preference tracking is most effective when the state is not only represented, but also searched over to choose the next action.

\begin{table}[t]
  \centering
  \small
  \setlength{\tabcolsep}{4pt}
  \caption{Prefix-level next-action evaluation. In Panel (a), \textit{w/Json Only} denotes \ours~without structured search. Panel (b) reports annotation reliability.}
  \label{tab:next_action_quality}
  \resizebox{0.48\textwidth}{!}{
  \begin{tabular}{lcccc}
    \toprule
    \multicolumn{5}{l}{\textbf{(a) Next-action quality}} \\
    Model & Acc. $\uparrow$ & Useful Ask $\uparrow$ & Redun. Q. $\downarrow$ & Prem. Rec. $\downarrow$ \\
    \midrule
    RA-CRS & 0.542 & 0.613 & 0.242 & 0.308 \\
    w/Json Only & 0.633 & 0.694 & 0.171 & 0.217 \\
    \ours & \textbf{0.758} & \textbf{0.806} & \textbf{0.089} & \textbf{0.117} \\
    \midrule
    \multicolumn{5}{l}{\textbf{(b) Annotation reliability}} \\
    Metric & \multicolumn{4}{c}{Agreement} \\
    \midrule
    Fleiss' Kappa on gold next action & \multicolumn{4}{c}{0.69} \\
    Krippendorff's alpha on error labels & \multicolumn{4}{c}{0.72} \\
    Human majority vs. LLM pre-label Macro-F1 & \multicolumn{4}{c}{0.77} \\
    \bottomrule
  \end{tabular}
  }
\end{table}

\subsection{Implementation of LLM Evaluation}\label{app_evaluator}

Prompt for the evaluation of \se~, \re~ and \ratio~ are in Figure \ref{fig:llm_re_se} and \ref{fig:llm_ratio}

\subsection{Implementation of Human Evaluation and Judge Reliability}
\label{app_human}

Inspired by ~\cite{zhang2024strength}, we conduct an interactive human evaluation on \ours~ under the ReDial dataset. We compare \ours~ with three competitive baselines, including PC-CRS, ChatCRS, and MACRS. Specifically, we hire 60 crowdworkers with diverse personas to interact with the four CRS models on movie recommendation tasks. To ensure objectivity, annotators are not allowed to communicate with each other during the evaluation process.

After the conversations, we collect dialogues for each CRS and evaluate their performance using Success Rate (SR), Recall@1, \se, and \re, as defined in Section~\ref{evaluation_protocol}. The final Fleiss' Kappa score reaches 0.71, indicating substantial agreement and supporting the reliability of the human verification process.

\noindent\textbf{Reliability of the LLM Judge.}
Prior work has shown that strong LLM judges can achieve relatively high agreement with human judgments under well-specified evaluation criteria, as evidenced by benchmarks such as MT-Bench and Chatbot Arena~\cite{zheng2023judging}. Although recent studies have identified limitations of LLM judges, such as positional bias and model favoritism~\cite{wang-etal-2024-large-language-models-fair}, these issues mainly arise in open-ended comparative evaluation settings. In contrast, our evaluator is only required to identify explicitly defined error types, making the judgment process more constrained. We therefore conduct an additional human validation study to examine its reliability. As summarized in Table~\ref{tab:judge_reliability}, the LLM judge achieves strong consistency with majority human annotations, suggesting that it can serve as a reliable and scalable proxy for our error-oriented evaluation. 
To further assess the reliability of our LLM-based evaluator, we randomly sample 50 dialogues generated by \ours~ on ReDial. We then ask 10 human annotators to label the utterance-level \se, \re, and the corresponding error ratios, following the same evaluation criteria used by the LLM evaluator. Human-human agreement is measured by Krippendorff's alpha, while LLM-human consistency is measured by Spearman's correlation over dialogue-level error ratios and macro-F1 against majority human labels at the utterance level. The Krippendorff's alpha scores are 0.74 and 0.68 for Elicitation and Exploitation, respectively, indicating substantial annotator agreement. The LLM judge also shows strong alignment with human majority labels, consistent with prior findings on LLM-based evaluators~\cite{liu2023g}. Detail results are shown in Table \ref{tab:judge_reliability}

\begin{table}[t]
\centering
\caption{Reliability validation of the LLM judge.}
\label{tab:judge_reliability}
\resizebox{0.49\textwidth}{!}{
\begin{tabular}{lccc}
\toprule
Metric & Human-Human & LLM-Human & LLM-Human \\
 & Krippendorff's $\alpha$ & Spearman's $\rho$ & Macro-F1 \\
\midrule
Overall & 0.70 & 0.72 & 0.76 \\
Elicitation / \se & 0.74 & 0.75 & 0.79 \\
Exploitation / \re & 0.68 & 0.69 & 0.73 \\
\bottomrule
\end{tabular}}
\end{table}

\subsection{Implementation of User Simulator \& Datasets}
\label{app_user}
Training and evaluating CRS with real user interactions can be impractically expensive at scale. To address this issue, we follow the previous LLM-based CRS\cite{qin2024beyond}\cite{fang2024multi} to employ ChatGPT\footnote{gpt-4o-mini-2024-07-18} as the user simulator.

For detail, we implement a goal-driven user simulator to interact with the CRSs under controlled and reproducible settings. At the beginning of each simulated dialogue, the user simulator is assigned a predefined persona and user preference, including preferred genres, actors, directors for movie recommendation and genres, writers for book recommendation. These preferences are sampled from a fixed pool of candidate attributes and are paraphrased into natural language to enhance realism. Specifically, we identify the 15 most common preference groups in Redial, 18 in OpendialKG for movie recommendation and 16 in OpendialKG for book recommendation. Additionally, in conjunction with 10 pre-defined diverse user personas, the evaluation process generates 150, 180 and 160 dialogues, respectively, for each dataset. Throughout the dialogue, the simulator generates context-aware responses based on its persona and the dialogue history. After receiving each recommendation from the system, the simulator successively executes three actions: (1) feeling inference, (2) insight generation, and (3) response generation.
Firstly, the user simulator internally evaluates the recommendation's alignment with its preferences through a brief, persona-consistent internal monologue. This internal state reflection guides the generation of follow-up intentions and subsequent user utterances. The simulator’s responses are generated using prompt-based calls to large language models, ensuring natural language variability, strict persona adherence, and coherence with the evolving conversation(See detailed prompts for feeling inference, insight generation and response generation in Tabel \ref{prompt:feeling_inference}, \ref{prompt:insight_generation}, \ref{prompt:response_generation} respectively). 

\subsection{Implementation of CRS-Simulator Interaction}
The interaction continues in a turn-by-turn manner until the simulator determines that a recommendation satisfies all criteria — namely, matching \textbf{all preferred genres }and \textbf{at least one preferred actor and director} — at which point the dialogue is concluded with a special \textbf{[END]} token. This simulation framework enables the systematic evaluation of recommender strategies under diverse user profiles and realistic conversational behaviors.

\subsection{Implementation of Memory Usage of \ours~}\label{memory}
The memory footprint of \ours~ is exceptionally low. The overhead introduced by the \ours's tree structure is minimal, primarily storing lightweight textual metadata like dialogue states and node scores, which avoids high-memory objects such as dense tensors. For instance, our GPT-4o-mini implementation keeps the total input text below 128k tokens (approximately 512KB using int32 encoding). Even with additional embeddings, the total memory usage remains around 375MB, a size that is well within the manageable limits of modern systems.

\begin{table}[ht]
\centering
\setlength{\abovecaptionskip}{5pt}
\setlength{\belowcaptionskip}{0pt}
\renewcommand{\arraystretch}{1.2}
\resizebox{0.45\textwidth}{!}{
\begin{tabular}{cccccc}
\toprule
$n(El)$ & $k(El)$ & $n(Ex)$ & Model & Time(s) & R@1 \\
\midrule
\multicolumn{3}{c}{--} & ChatCRS   & 8 & 0.422 \\
\multicolumn{3}{c}{--} & PC-CRS    & 13 & 0.378 \\
\multicolumn{3}{c}{--} & RA-CRS    & 19 & 0.267 \\
\multicolumn{3}{c}{--} & MACRS     & 18 & 0.289 \\
\multicolumn{3}{c}{--} & InterCRS  & 5 & 0.244 \\ \midrule
 3  & 3 & 3  & \ours & 23 & 0.489 \\
\midrule
5  & 3 & 3  & \ours & 34 & 0.578 \\
10 & 3 & 3  & \ours & 75 & 0.600 \\
\midrule
\rowcolor{gray!20} 3  & 2 & 3  & \ours & 14 & 0.422 \\
3  & 1 & 3  & \ours & 10 & 0.333 \\
\midrule
\rowcolor{gray!20} 3  & 2 & 5 & \ours & 16 & 0.444 \\
3  & 3 & 5  & \ours & 25 & 0.533 \\
3  & 3 & 10 & \ours & 30 & 0.489 \\
\midrule
\rowcolor{gray!20} \multicolumn{3}{c}{--} & \ours(EA)   & 9 & 0.467 \\

\bottomrule
\end{tabular}
}
\caption{Time efficiency analysis. $n$ and $k$ represent iteration steps and simulation depth. (El) and (Ex) denote parameters for El- and ExNodes, respectively. \ours~excels in high-stakes environments requiring maximum effectiveness, whereas \ours(EA) is tailored for scenarios where low latency is critical.}
\label{tab:dreams_hyperparameter}
\vspace{-3mm}
\end{table}

\section{Additional Analysis}
\subsection{Practicality and Efficiency Analysis}
\label{app:efficiency}

\noindent\textbf{Efficiency Optimization via Experience-augmentation}. Table \ref{tab:dreams_hyperparameter}  indicates that the time efficiency of vanilla \ours\ is comparable to existing CRS methods. However, a common concern with tree-based planning is latency. Inspired by \citet{gu-etal-2025-toward, ren2025sigmarefininglargelanguage}, we leverage the MCTS trees from the training phase as historical experience. This allows us to bypass the time-consuming tree search during deployment. In particular, we introduce \textbf{\ours(EA)}, a \underline{E}xperience-\underline{A}ugmented version. Crucially, we construct an Experience Knowledge Base (EKB) using 50 held-out user simulators to interact with \ours~ under the settings of $n(El)=5$, $k(El)=3$, and $n(Ex)=3$. We record tuples $\langle \mathcal{D}, S, a, r, q', R \rangle$ that represent the dialogue context, structured state, action, decision reason, refined query, and final reward, respectively. During the inference phase, prior to generating a response, we retrieve relevant priors by calculating the semantic similarity between the current structured state and the stored experience states in the EKB. Unlike standard text-based retrieval, matching on structured states ($S$) ensures that the retrieved contrastive pair (one successful and one failed exemplar) shares structural isomorphism with the current situation, effectively filtering out contextual noise. By instructing the model to ``\textit{follow the successful pattern and avoid specific errors}'', this mechanism acts as a warm-start that wisely guide the strategic planning of CRS. As shown in Table \ref{tab:dreams_hyperparameter}, the average inference time drops to 9s, while Recall@1 improves to 0.467, confirming that \ours(EA) is an efficient and effective solution for real-time applications.

\begin{table}[ht]
  \centering
  \setlength{\abovecaptionskip}{5pt}
\setlength{\belowcaptionskip}{0pt}
  \resizebox{0.5\textwidth}{!}{%
  \begin{tabular}{l ccc ccc}
    \toprule
    Backbone & \multicolumn{3}{c}{Gemini-2.5-flash (1M)} & \multicolumn{3}{c}{GPT-4o-mini (128K)} \\
    \cmidrule(lr){2-4} \cmidrule(lr){5-7}
    & SR & $FGE^2$ & $CGE^2$ & SR & $FGE^2$ & $CGE^2$ \\
    \midrule
    \ours~ (Ours) & \textbf{0.587} & \textbf{0.180} & \textbf{0.277} & \textbf{0.560} & \textbf{0.100} & \textbf{0.289} \\
    InterCRS (free-form) & 0.347 & 0.347 & 0.503 & 0.313 & 0.360 & 0.570 \\
    RA-CRS (structured) & 0.387 & 0.340 & 0.449 & 0.333 & 0.333 & 0.461 \\
    \bottomrule
  \end{tabular}
  }
  \caption{Performance on the Redial dataset across LLM backbones with 128K/1M token limits.}
  \label{tab:gemini}
\end{table}

\noindent\textbf{Hyperparameter and Computational Overhead.} \ours~has three hyperparameters: $n(El)$, $k(El)$ and $n(Ex)$. Results in Table~\ref{tab:dreams_hyperparameter} show that the model's performance improves significantly when the number of elicitation iterations ($n(El)$) is increased to 5 and the exploitation iterations ($n(Ex)$) contains at least 3. However, further increases lead to diminishing returns, with marginal gains or even slight performance degradation. 
Increasing iterations and simulation depth enhances accuracy but reduces time efficiency. We employ multi-threading to parallelize computation in both Node types, ensuring \ours~achieves comparable time efficiency and better performance. For example, under specific configuration, the computational expense of \ours~(16s/turn) is commensurate with that of MACRS (18s/turn), yet it demonstrates a markedly superior performance metric (0.444 VS. 0.289). On this turn, \ours(EA) achieves a remarkable balance: it reduces the average inference time to 9s while maintaining a competitive Recall@1 of 0.467, significantly outperforming standard baselines like MACRS and PC-CRS in both speed and accuracy. Thus, \ours~excels in high-stakes environments requiring maximum effectiveness, whereas \ours(EA) is tailored for real-time scenarios where low latency is critical.

Notably, both the standard \ours~and \ours(EA) introduce no substantial memory overhead. This is because the tree structure and the Experience Knowledge Base predominantly store lightweight textual metadata, such as dialogue states, node scores, and reasoning tuples, rather than high-memory objects like dense tensors. Even with the addition of historical experience entries in \ours(EA) and the embeddings used during retrieval, the total memory usage remains approximately 560MB. This ensures that the system remains well within the limits of modern hardware while significantly outperforming baselines in real-time scenarios.

\subsection{Case Study}
\label{app:case_study}
\begin{figure*}[t]
    \centering
    \includegraphics[width=0.9\textwidth]{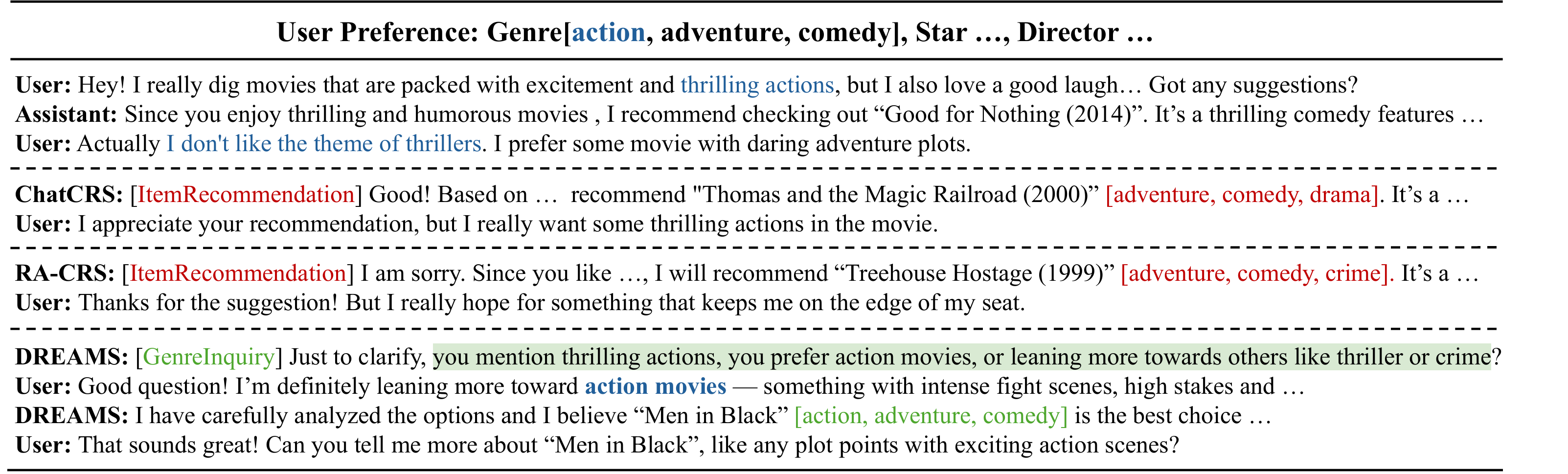}
    \setlength{\abovecaptionskip}{5pt}
\setlength{\belowcaptionskip}{0pt}
    \caption{Case study illustration. The user's initial raw preference, containing the subtly misleading phrase (blue, non-bold text), often causes the CRSs to misinterpret the intent (e.g., mistaking "thrilling actions" (target preference is Action movie) for the Thriller genre, leading to an error spiral where baselines repeatedly act on the flawed understanding. \ours~identifies the critical error using structured context modeling and successfully navigates to the optimal path (green, non-bold text).
    }
    \label{fig:case_study}
    \vspace{-3mm}
\end{figure*}

Figure \ref{fig:case_study} and Table \ref{tab:mcts_trajectory} provide a concrete mechanism-level example. In the case study, the user's phrase ``thrilling action'' creates an ambiguous signal: the user intends action movies, but the surface phrase can be misread as the thriller genre. ChatCRS and RA-CRS recommend too early and enter an error spiral based on this flawed interpretation. \ours~ instead selects GenreInquiry and asks whether the user means action movies or thriller/crime movies. Table \ref{tab:mcts_trajectory} explains why this decision emerges. The LLM prior assigns the highest score to ItemRecommendation (0.44), since the user appears to request a recommendation. However, MCTS rollouts assign negative reward to immediate recommendation and high reward to GenreInquiry. As a result, GenreInquiry obtains the highest visit ratio (0.667) and final score (0.524), so \ours~ clarifies the preference before retrieval. This case shows that structured search changes decision making from following the most likely immediate response to evaluating which action best improves preference tracking.
Appendix \ref{app:next_action_quality} further evaluates next-action quality on 120 annotated Redial prefixes. \ours~ improves Next-Action Accuracy from 0.633 for w/Json Only to 0.758 and reduces Premature Recommendation Rate from 0.217 to 0.117, confirming that structured search makes the preference state actionable.

\begin{table}[t]
  \centering
  \small
  \caption{MCTS trajectory for the ambiguous-preference prefix in Figure~\ref{fig:case_study}. The LLM prior favors immediate recommendation, but rollouts select GenreInquiry after evaluating downstream preference-tracking utility.}
  \label{tab:mcts_trajectory}
  \resizebox{0.5\textwidth}{!}{
  \begin{tabular}{lccccc}
    \toprule
    Cand. Action & LLM prior & Visit ratio & Reward & Score & Decision \\
    \midrule
    GenreInq. & 0.35 & \textbf{0.667} & \textbf{10.2} & \textbf{0.524} & selected \\
    ActorInq. & -- & -- & -- & -- & not selected \\
    DirectorInq. & 0.21 & 0 & -4.9 & 0.098 & not selected \\
    ItemRec. & \textbf{0.44} & 0.333 & -4.6 & 0.381 & not selected \\
    ItemExp. & -- & -- & -- & -- & not selected \\
    \bottomrule
  \end{tabular}
  }
\end{table}

\begin{table*}[t]
    \centering
    \renewcommand\arraystretch{1.2}
    \resizebox{1.0\textwidth}{!}{%
    \begin{tabular}{p{2cm}p{6cm}p{18cm}}
        \toprule
        \textbf{Iteration ID} &
          \textbf{Selection, Expand and Reward} &
          \textbf{Dialogue simulation}  \\ \midrule
        1 &
          \begin{tabular}[c]{@{}p{6cm}@{}}\textbf{Selection}: no children\\\\
          \\\textbf{Expand}: [ItemRecommendation]
          
          \\\\
          
          \textbf{Reward}: 2.5
          
          \end{tabular} &
          \begin{tabular}[c]{@{}p{18cm}@{}}
            sim\_assistant\_response:(ItemRecommendation)\\
            Got it! If you're looking for an exciting adventure without the thrilling aspect, I recommend " Jaws " (1975), directed by Steven Spielberg. It combines thrilling, and adventure themes of faith, honor, and the complexities of war, making it a thrilling watch that aligns well with your preference. \\
            sim\_user\_feedback: \\
            I've seen " Jaws " before, and while I appreciate Steven Spielberg 's work, it didn't quite resonate with me. Do you have any other recommendations, maybe something with a bit more action?\\
            sim\_assistant\_response: (ItemRecommendation)\\
            Thanks for the feedback! If you’re looking for something with more action, you might enjoy “The Last Samurai” (2003) — it’s rich in intense battle scenes and emotional depth. If you’re open to something more recent, “13 Assassins” (2010) offers gripping samurai action with stunning choreography. Would you prefer historical settings, or are you open to modern or sci-fi action as well?\\
            sim\_user\_feedback: \\I appreciate the effort, but none of those really catch my interest. Can you suggest something else, maybe focusing more on the actors I Like?\\
            sim\_assistant\_response: \\
            {[}FailureReflection{]} “The user rejected my recommendation and mentioned that he wanted some recommendations related to his favorite stars or directors. But I don’t know who his favorite stars or directors are. So I should first ask him which stars or directors he likes.”\\
            {[}StarInquiry{]} That’s ok. Can you share me with some of your preferred stars and directoes?\\
            sim\_user\_feedback: \\
            Enjoy to share that I am a fan of Jackie Chan, and I love the style of Barry Sonnenfeld\\
          \end{tabular}

          \\ \hline
        2 &
          \begin{tabular}[c]{@{}p{6cm}@{}}
          \textbf{Selection}: not\_fully\_expanded\\\\
          \\\textbf{Expand}: ["FailureReflection"]\\
          {[}Failure Reflection{]} “The user has explicitly rejected the recommendation of 'Good for Nothing (2014)' due to its theme, indicating a misunderstanding of their preferences. Therefore, the most appropriate action is to acknowledge this misunderstanding and reset the conversation by revisiting preference inquiries”
          \\\\
          
          \textbf{Reward}: 20.1
          
          \end{tabular} &
          \begin{tabular}[c]{@{}p{18cm}@{}}
            sim\_assistant\_response: {[}GenreInquiry{]}\\
            To clarify, when you mention "thrilling actions," do you mean action-oriented films specifically, or are you referring more broadly to genres such as thrillers or crime? \\
            sim\_user\_feedback: \\
            I’d say I’m leaning more toward action movies — ones that feature intense fight scenes, high stakes, and fast-paced plots. Do you have any recommendations in that vein?\\
            sim\_assistant\_response: {[}ItemRecommendation{]}\\
            Got it! Based on your interest in high-stakes, adrenaline-pumping action, daring adventures and something that can make you good laugh, I will recommend “Men in Black”. It’s an action movie with adventurous and humorous elements, which meets all your preference.\\
            sim\_user\_feedback: \\
            That sounds awesome. Can you tell me morec about “Men in Black” like it features which stars and who directs it?\\
            sim\_assistant\_response: {[}ItemExplanation{]}\\
            Absolutely! “Men in Black” is a classic action-comedy film directed by Barry Sonnenfeld. The film stars Will Smith as Agent J and Tommy Lee Jones as Agent K, two secret agents working for a covert government organization that monitors extraterrestrial activity on Earth.\\
            Based on the comic book series by Lowell Cunningham, the story follows Agent J as he is recruited into the Men in Black and partners with the veteran Agent K to investigate alien-related incidents. Together, they uncover a plot that threatens both Earth and the intergalactic balance, all while navigating humorous and high-stakes encounters with aliens living secretly among humans.\\
            sim\_user\_feedback: \\That definitely matched all my preference! I will give it a try tonight. Thank you {[}END{]}
          \end{tabular}
          \\ \hline
        3 &
          \begin{tabular}[c]{@{}p{6cm}@{}}
          \textbf{Selection}: {[}FailureReflection{]}\\\\
          \\\textbf{Expand}: fully\_expanded
          \\\\
          
          \textbf{Reward}: 10.5
          
          \end{tabular} &
          \begin{tabular}[c]{@{}p{18cm}@{}}
            sim\_assistant\_response (FailureReflection):\\
            "I see that my last recommendation might not have aligned well with your preferences — I may have focused too much on suspense rather than fun and excitement. Let me ask a quick question to get a clearer sense of what you're looking for."\\
            sim\_user\_feedback:\\
            "Thanks. Yeah, I'm more into movies that are fun to watch — action, adventure, maybe with a bit of comedy thrown in. Not too serious or dark."\\
            sim\_assistant\_response (GenreInquiry):\\
            "Got it! Just to be sure, do you prefer modern action films or do you enjoy older classics too?"\\
            sim\_user\_feedback:\\
            "I’m open to both, as long as they’re fast-paced and entertaining."\\
            sim\_assistant\_response (ItemRecommendation):\\
            "In that case, you might enjoy “The Man from U.N.C.L.E.” (2015). It’s stylish, full of clever action sequences, and has a good dose of humor and charm."\\
            sim\_user\_feedback:\\
            "Hmm, I’ve heard about it. Sounds decent, but I’ll have to see a bit more before deciding."\\

          \end{tabular}
          \\ \hline
        4 &\begin{tabular}[c]{@{}p{24cm}@{}}
          \textbf{Possible actions}:\\
          1. FailureReflection: mcts-score: 0.66666667, llm-score: 0.44444444\\
          2. ItemRecommendation: mcts-score: 0.33333333, llm-score: 0.55555556\\
          \textbf{Combined Scores}:\\FailureReflection: 0.6, ItemRecommendation: 0.4\\
          \textbf{Best\_action}: FailureReflection\\
          \textbf{User}: {[}FailureReflection{]}
        \end{tabular} 
            
           \\ \bottomrule
    \end{tabular}%
}
\caption{}
\label{tab:sim}
\end{table*}

\section{Prompts}\label{prompts}
Here are detailed prompts for aforementioned implementations.

\begin{table*}[h]
\centering
\begin{tabular}{p{12cm}}
\toprule
\textbf{Prompt Template for Feeling Inference} \\
\midrule
The Seeker notes how he feels to himself in one sentence. \\
What aspects of the recommended movies meet your preferences? What aspects may not? \\
What do you think of the performance of this recommender? \\
What would the Seeker think to himself? What would his internal monologue be? \\
\textit{The response should be short (as most internal thinking is short) and strictly follow your Seeker persona.} \\
\textit{Do not include any other text than the Seeker's thoughts.} \\
\textit{Respond in the first-person voice (use "I") and speaking style of Seeker. Pretend to be Seeker!} \\
\bottomrule
\end{tabular}
\caption{Prompt for Feeling Inference of User Simulator.}
\label{prompt:feeling_inference}
\end{table*}

\begin{table*}[h]
\centering
\begin{tabular}{p{12cm}}
\toprule
\textbf{Prompt Template for Insight Generation } \\
\midrule
Here is your feelings about the last sentence: \texttt{<seeker\_feelings>} \\
What's the insight and next plan of the Seeker in one sentence based on the information above? \\
What would the Seeker think to himself? What would his internal monologue be? \\
\textit{The response should be short (as most internal thinking is short).} \\
\textit{Do not include any other text than the Seeker's thoughts.} \\
\textit{Respond in the first-person voice (use "I") and speaking style of Seeker. Pretend to be Seeker!} \\
\bottomrule
\end{tabular}
\caption{Prompt for Insight Generation of User Simulator.}
\label{prompt:insight_generation}
\end{table*}

\begin{table*}[h]
\centering
\begin{tabular}{p{12cm}}
\toprule
\textbf{Prompt Template for Response Generation} \\
\midrule
Here is your feelings about the last sentence: \texttt{<seeker\_feelings>} \\
Here is your insights about what to do next: \texttt{<seeker\_insights>} \\
What does the Seeker say next? \\
\textit{Keep your response brief. Use casual language and vary your wording.} \\
\textit{Make sure your response matches your Seeker persona, your preferred attributes, and your conversation context.} \\
\textit{Do not include your feelings into the response to the recommender!} \\
\textit{Respond in the first-person voice (use "I" instead of "Seeker", and "you" instead of "recommender"). Pretend to be Seeker!} \\
\bottomrule
\end{tabular}
\caption{Prompt for Response Generation of User Simulator.}
\label{prompt:response_generation}
\end{table*}

\begin{figure*}[ht]
\begin{tcolorbox}[left=1mm,right=1mm,top=0.mm, bottom=0mm,colback=white]
\begin{lstlisting}[style=demo]
You are an experienced dialogue reviewer. Please help me analyze the following dialogue history for the presence of "invalid recommendations""invalid questions""repeated questions" and "retriever error".
    The definitions of these terms are as follows:
    "FGE2(Fine-Grained Elicitation Error)" refers to a situation where the assistant, after the user rejects a recommendation in a previous round, fails to reflect on the feedback and does not ask the appropriate questions, but proceeds to make another recommendation directly or ask some unmatched issues(e.g. The user says he dislikes the recommended movie because it's a horror movie, the assistant shouldn't ask the user about his preferred stars).
    
    "PE2(Preference Exploitation Error)" refers to a situation where the assistant has already figured out explicit user preference about all the attribute(genre, actor and director), while the "retrieval result" is still false.
Dialogue History
##########

Response in the following JSON format:
{"FGE2": <bool>, "PE2": <bool>}'''
\end{lstlisting}
\end{tcolorbox}
\vspace{-3.5mm}
\caption{LLM Evaluation Prompt for \re and \se}
\label{fig:llm_re_se}
\end{figure*}

\begin{figure*}[ht]
\begin{tcolorbox}[left=1mm,right=1mm,top=0.mm, bottom=0mm,colback=white]
\begin{lstlisting}[style=demo]
You are tasked with analyzing a conversation history stored in a JSON file that contains dialogues between a user and an assistant. 
    Each dialogue turn by the assistant includes information on whether the recommendation system's retrieval results were correct or incorrect (recorded in the 'retrieval results' field). 
    Your job is as follows:
    1. Intent Classification. For each statement made by the assistant, determine whether its intent is:
    Ask : When the assistant is asking a question.
    Recommend : When the assistant is making a recommendation.
    Others : When the statement does not fall into either of the above categories.
    2. State Classification. Check if the user has already explicitly expressed all his preference about genre, actor and director
    3. Question Checking. When the assistant asks a question, check if this question conforms to the current context. For example, if the user has not revealed their favorite stars before, asking questions related to stars is an appropriate question. For another example, if the user claims that he doesn't like the director, the assistant should ask for his preferred directors. But if the assistant asks about the preferred stars, it's a invalid question that doesn't confirm the context.
    4. Relevance Evaluation. Based on the 'retrieval results' field, classify each instance into one of the following categories:
    Ask-True : The assistant takes action 'ask' when the 'retrieval results' is 'True'.
    Ask-False : The assistant takes action 'ask' when the 'retrieval results' is 'False'.
    Recommend-True : The assistant takes action 'recommend' when the 'retrieval results' is 'True'.
    Recommend-False : The assistant takes action 'recommend' when the 'retrieval results' is 'False'.
    3. Output Requirements : Generate a summary count of the following categories:
    Total instances of Ask-True
    Total instances of Ask-False
    Total instances of Recommend-True
    Total instances of Recommend-False
    4. Attention! 
    Please do not label the user's statements; focus only on the statements made by the assistant.
    If you determine that the intent of the assistant in a particular dialogue turn is neither "ask" nor "recommend" (e.g., it could be "greetings""explanations" or similar), do not include it in any metrics related to Ask-True, Ask-False, Recommend-True, or Recommend-False.
    
Dialogue History
##########

Response in the following JSON format:
{"Ask-True": <int>, "Ask-False": <int>, "Recommend-True": <int>, "Recommend-False": <int>, "Invalid-Recommendation": <int>, "Invalid-Question": <int>}
\end{lstlisting}
\end{tcolorbox}
\vspace{-3.5mm}
\caption{LLM Evaluation Prompt for \ratio}
\label{fig:llm_ratio}
\end{figure*}

\begin{figure*}[ht]
\begin{tcolorbox}[left=1mm,right=1mm,top=0.mm, bottom=0mm,colback=white]
\begin{lstlisting}[style=demo]
You are an AI assistant helping to determine the best action for a conversational movie recommender system.
Based on the current conversation state(mainly focusing on the user preference, user attitude, and recent actions), rank the following actions from most to least appropriate:
- GenreInquiry: Ask about the user's preferred movie genres
- ActorInquiry: Ask about the user's preferred actors
- DirectorInquiry: Ask about the user's preferred directors
- ItemRecommendation: When explicitly knowing all of the user preference(genre, actor, director), retrieve for the movie item and recommend it to the user.
- ItemExplanation: After you recommend a movie, if the user is interested in the recommended movie and ask for details, provide details about the recommended movie to the user
- FailureReflection: When your recommendation is rejected by the user, reflect on the fundamental reasons for being rejected and take corresponding measures.


Here are some empirical rules:
1.	When the attribute of user preference in the state is empty, start with GenreInquiry, then narrowing down with ActorInquiry or DirectorInquiry.
2.  Attention! Unless the user has already expressed their preference about all the attributes(genres, actors, directors), you should consider the user preference as not clear.
3.  After you get the explicit user preference about all the attributes(genres, actors, directors), you should conduct ItemRecommendation.
4.	Use FailureReflection for recovery: If recommendations fail(user attitude is rejected), reset by revisiting preference inquiries.
5.	Enhance recommendations with explanations: After Recommendation, provide ItemExplanation to improve user trust.
6.  Do not proceed with ItemExplanation before ItemRecommendation has been conducted.
7.  If the user continuously ask about the details of the movie in successive rounds, you can continuously adopt "ItemExplanation" strategie.

Here are some examples of optimal decision paths:
Example 1:
GenreInquiry -> ActorInquiry -> DirectorInquiry -> ItemRecommendation -> ItemExplanation
Example 2:
GenreInquiry -> ActorInquiry -> ItemRecommendation -> ItemExplanation -> FailureReflection -> DirectorInquiry -> ItemRecommendation -> ItemExplanation


Also, you should give your reason for the score of each action.
Return a JSON object with each action and a score from 0-10 where 10 is most appropriate, and a reason for the score:
{
    "GenreInquiry": score,
    "ActorInquiry": score,
    "DirectorInquiry": score,
    "ItemRecommendation": score
    "ItemExplanation": score,
    "FailureReflection": score,
    "reason": reason
}
\end{lstlisting}
\end{tcolorbox}
\vspace{-3.5mm}
\caption{LLM Scoring Prompt of \ours}
\label{fig:llm_score}
\end{figure*}

\end{document}